\documentstyle[12pt]{article}

\makeatletter
\def\section{\@startsection{section}{1}{\z@}{-3.5ex plus -1ex minus
 -.2ex}{2.3ex plus .2ex}{\large\bf}}
\def\subsection{\@startsection{subsection}{2}{\z@}{-3.25ex plus -1ex minus
 -.2ex}{1.5ex plus .2ex}{\normalsize\bf}}
\makeatother

\makeatletter
\@addtoreset{equation}{section}

\makeatother

\begin{document}

\vspace{8mm}

\begin{center}

{\Large \bf Partition Function of a Quadratic Functional \\
and \\ Semiclassical Approximation for Witten's 3-Manifold
Invariant\footnote{Work supported by FORBAIRT Scientific Research
Program SC/94/218} } \\

\vspace{12mm}
{\large David H.~Adams\footnote{Supported by Trinity College Postgraduate
Award, FORBAIRT Basic Research Award,
Danish Research Academy and Knud Hojgaards Fond}
and Siddhartha Sen }

\vspace{4mm}
School of Mathematics, Trinity College, Dublin 2, Ireland. \\
\vspace{1ex}
email: dadams,sen@maths.tcd.ie   \\
\end{center}

\vspace{4mm}

\vspace{4mm}

\begin{abstract}
An extension of the method and results of A.~Schwarz for evaluating
the partition function of a quadratic functional is presented.
This enables the partition functions to be evaluated for a wide class
of quadratic functionals of interest in topological quantum field theory,
for which no method has previously
been available. In particular it enables
the partition functions appearing in the semiclassical approximation for the
Witten-invariant to be evaluated in the most general case.
The resulting $k-$dependence is precisely that
conjectured by D.~Freed and R.~Gompf.
\end{abstract}

\newpage

\section{Introduction and summary}

Partition functions of quadratic functionals (of the form
(\ref{1.1}) below) are
important when studying the topology of manifolds via the methods of
quantum field theory (QFT). In 1978 A.~Schwarz \cite{S1} \cite{S} showed that
the Ray-Singer analytic torsion \cite{RS1}, a topological invariant, could be
obtained within a QFT framework as the partition function of a certain
quadratic functional. Since then it has been realised that other topological
formulae and results can be obtained within a QFT framework (see \cite
{BirRak} and  \cite{BlauThom} for reviews); this often involves evaluating
partition functions of quadratic functionals.

These partition functions play an important role in recent interaction
between QFT and 3-dimensional topology. In 1988 E.~Witten \cite{W} constructed
a powerful new invariant of 3-manifolds within the framework of Chern-Simons
gauge theory, a topological QFT\footnote{It was the subsequent work of
Reshetikin-Turaev \cite{RT} and K.~Walker \cite{Walker} that showed that
this invariant really is a well-defined topological invariant.}.
Since its discovery the Witten-invariant has attracted considerable interest
and has been explicitly calculated for a number of 3-manifolds and gauge
groups, see for example \cite{KirMel},\cite{DiWi},\cite{J},\cite{Roz},
\cite{RaSe} and the references therein.
As well as
providing a new tool for studying the topology of 3-manifolds Witten's work
opens up a possibility for testing predictions of the standard methods of
QFT. This is interesting not only from a physics point of view, but from a
mathematics point of view as well: If the predictions of QFT hold in this
case then, as we will discuss at the end of this \S,
it indicates deep and hitherto
unexplored relationships between different areas of mathematics. The
Witten-invariant associates to a 3-manifold $M$ and gauge group $G$ a function
$Z_W(k)$ of a parameter $k\in\bf{Z}\,$,
and the most basic prediction of QFT is
an expression for $Z_W(k)$ in the limit $k\to\infty\,$, known as the
semiclassical approximation. This expression,
(\ref{1.14}) below, is essentially
a sum over partition functions of quadratic functionals.

The large$-k$ limit of $Z_W(k)$ has been explicitly calculated for a number of
3-manifolds (with $G=SU(2)\,$) in a program initiated by D.~Freed and
R.~Gompf in \cite{FG} \cite{FG1}\footnote{In this work the large$-k$ limit of
$Z_W(k)$ was calculated numerically for a number of lens spaces and
Brieskorn spheres. It was later calculated analytically by L.~Jeffrey \cite{J}
for lens spaces and torus bundles over the circle,
and by L.~Rozansky \cite{Roz}
for Seifert manifolds.}. However, it has not been possible to make
a complete comparison with the semiclassical approximation predicted by QFT
because no method has existed for evaluating the partition functions appearing
in this expression in general. In a very restricted case it has been possible
to partially evaluate the partition functions, this was done by Witten in
\cite[\S2]{W}.
The expression obtained is independent of a choice of metric used to
construct it, except for a phase factor. Witten showed that the
metric-dependence of the phase factor can be cancelled by adding a
"geometric counterterm", i.e. functional of the metric to the
phase of each partition function.

Based on the expression obtained in this restricted case and
on the results of their numerical calculations of the large$-k$ limit of
$Z_W(k)\,$, Freed and Gompf conjectured an expression for the partition
functions in the semiclassical approximation in the general case, which leads
to agreement with the large$-k$ limit of $Z_W(k)\,$\footnote{The agreement is
modulo a discrepancy in the overall numerical factor. This discrepancy was
removed in the subsequent works \cite{J} and \cite{Roz}
where the conjectured expression of Freed and Gompf was refined.}
In particular they conjectured the form of the dependence of the partition
functions on the parameter $k\,$; this is not obtained in Witten's partial
evaluation of the partition functions in the restricted case.

In this paper we give a method for evaluating the partition function for a
wide class of quadratic functionals for which no method has previously
existed. This includes all the partition functions appearing in the
semiclassical approximation for the Witten-invariant. The expression we obtain
in this case agrees with that conjectured by Freed and Gompf
except for a discrepancy in the phase and overall numerical factor.
In particular the conjectured $k-$dependence is obtained.
The phase differs from the conjectured expression by Witten's geometric
counterterm; as before this must be added by hand to obtain a topological
invariant which agrees with the large$-k$ limit of $Z_W(k)\,$.
We discuss how the discrepancy in the numerical factor can be
understood from recent work by the first author combined with the results of
Rozansky \cite{Roz}. Thus we resolve a number of the ``deeper theoretical
mysteries'' posed as problems for the reader by Freed and Gompf in the
conclusion of their paper \cite{FG}.

Our method for evaluating partition functions of quadratic functionals is an
extension and refinement of the method of A.~Schwarz \cite{S1} \cite{S}.
We prove that certain invariance-properties of the partition function derived
by Schwarz in a restricted case continue to hold in the general case which we
consider. This leads to our expression for the partition function being a
topological invariant for a wider class of quadratic functionals. For example
the Ray-Singer torsion ``as a function of the cohomology'' \cite{RS2} is
obtained as the partition function of a quadratic functional. We also derive
a new invariance-property of the partition function which relates the previous
ones. We derive expressions for
the phase of the partition function and the dependence of the
partition function on a complex-valued scaling parameter multiplying the
quadratic functional. This was not considered in Schwarz's work\footnote{The
expression derived by Schwarz for the partition function is really
the modulus of the partition function.} and is necessary for evaluating the
partition functions in the semiclassical approximation for the
Witten-invariant.

We apply our metric-independence results to derive a result concerning the
usual Ray-Singer torsion of a flat connection. When the cohomology of the
connection is non-vanishing the torsion is metric-dependent, however we show
that in certain cases the metric-dependence factors out in a simple way
as a power of the volume of the manifold
to give a topological invariant.

To motivate the method (and also to demonstrate it in a simple context) we
devote a section of this paper (\S2) to studying the partition function of
a quadratic functional on a finite-dimensional vectorspace. In this case it
turns out that the method can be motivated by requiring that a
symmetry-property of the partition function of a non-degenerate quadratic
functional continues to hold when the functional is degenerate.

The results we obtain in this finite-dimensional case may be of independent
interest in algebraic geometry. It turns out that the partition function
is essentially the determinant of a complex of linear maps between
vectorspaces. Given inner products in these vectorspaces, and in the
cohomology spaces of the complex, the norm of the determinant is a
well-defined number. We derive formulae expressing the change in (the
norm of) the determinant under changes in the inner products and changes in
the maps between the vectorspaces. These formulae hold for an arbitrary
(finite) complex of linear maps and are generalisations of formulae for
ordinary determinants.
They are simple to write down but non-trivial to prove.
{}From them we find classes of changes under which the determinant is
invariant.

We now briefly describe what the method involves (the details will be given
in \S2 and \S3); this will enable us to describe
our results more concretely. The partition functions of quadratic functionals
which we consider are formal integrals over spaces of functions, or more
generally sections in a vectorbundle over a manifold $M\,$. They have the
form
\begin{equation}
Z(\beta)=\int_{\Gamma}\,{\cal D\/}\omega \,e^{-{\beta}S(\omega)}
\label{1.1}
\end{equation}
where $S(\omega)$ is a real-valued quadratic functional on the space
$\Gamma$ of sections $\omega\,$
(i.e. $S(\omega)=F(\omega,\omega)$ where $F(\omega,
\upsilon)$ is a real-valued bilinear functional on $\Gamma\,$) and $\beta$
is a complex-valued scaling parameter. (Typically $\beta$ is either real or
purely imaginary; sometimes $\beta$ is taken to be a constant equal to 1
or $-i\,$).
The  expression (\ref{1.1}) is
mathematically ill-defined in general. This is due firstly to the fact that
the integration is over an infinite-dimensional vectorspace $\Gamma\,$
(when $dimM>0\,$), and secondly, even when $\Gamma$
is finite-dimensional (i.e. when $dimM=0\,$) (\ref{1.1}) diverges unless
${\beta}S(\omega)$ is positive and non-degenerate.
The method we describe for evaluating and normalising
(\ref{1.1}) involves formal
manipulations with mathematically ill-defined quantities. It leads to a
well-defined finite expression for $Z(\beta)$ in the cases with which we will
be concerned, and the results we subsequently derive for this expression
are all mathematically rigorous.

A choice of inner product
$\langle\cdot\,,\cdot\rangle$ in $\Gamma$ determines (formally) an
integration measure ${\cal D\/}\omega$ and allows us to write
$S(\omega)={\langle}\omega,T\omega\rangle$
where $T$ is a uniquely determined selfadjoint map.
Decomposing $\Gamma=ker(T){\oplus}ker(T)^{\perp}\,,\;\omega=
(\omega_1,\omega_2)$ and ${\cal D\/}\omega={\cal D\/}\omega_1{\cal D\/}
\omega_2$ we can formally write
\begin{eqnarray}
Z(\beta)
&=&\int_{ker(T)}{\cal D\/}\omega_1\,\biggl(\,\int_{ker(T)^{\perp}}{\cal D\/}
\omega_2\,e^{-\beta\langle(\omega_1,
\omega_2),T(\omega_1,\omega_2)\rangle}\biggr)\nonumber \\
&=&V(ker(T))\,\int_{ker(T)^{\perp}}{\cal D\/}\omega_2\,
e^{-\langle\omega_2,\beta\widetilde{T}\omega_2\rangle}
\label{1.2}
\end{eqnarray}
where $\widetilde{T}\,\colon\,ker(T)^{\perp}\stackrel{\simeq}{\longrightarrow}
ker(T)^{\perp}$ is the restriction of $T$ to $ker(T)^{\perp}$ and
$V(ker(T))$ is the (divergent) volume of $ker(T)\,$.
For simplicity we assume for the moment that $\beta\in\bf{R_+}$ and
$S(\omega)$ is positive. Then (\ref{1.2}) can be formally evaluated to get
\begin{equation}
Z(\beta)=\pi^{\zeta/2}\,\beta^{-\zeta/2}\,det(\widetilde{T})^{-1/2}
\,V(ker(T)) \label{1.3}
\end{equation}
where $\zeta=dim\Gamma-dim(ker(T))\,$\footnote{The numerical factor
$\pi^{\zeta/2}$ in
(\ref{1.3}) should not be discarded because it is relevant
when comparing the semiclassical approximation with the large$-k$ limit
of the Witten-invariant.}. All the factors in the formal
expression
(\ref{1.3}) are divergent in general; however in the cases with which
we will be concerned $\zeta$ and  $det(\widetilde{T})$ can be regularised
via zeta-function regularisation techniques to obtain finite expressions.
Then, if $S$ is non-degenerate,
$ker(T)=0$ and (\ref{1.3}) gives a finite
expression for the partition function, which depends on the choice of
inner product $\langle\cdot\,,\cdot\rangle$ in $\Gamma\,$.
When $S(\omega)$ is degenerate the partition function diverges due to the
divergent volume $V(ker(T))$ in
(\ref{1.3}) and must be ``normalised'' to
obtain a finite expression. The simplest way to do this is to simply
divide out (i.e. discard) the factor $V(ker(T))$ in
(\ref{1.3}). However, this
is not compatible with the usual QFT procedure of
Faddeev and Popov. Also, as we will see in \S2, it does not preserve a
symmetry-property of the partition function of non-degenerate
functionals under changes in the inner product $\langle\cdot\,,
\cdot\rangle$ in $\Gamma\,$.
The method for evaluating the partition function in the degenerate case
requires the functional $S$ to have an additional structure associated
with it, namely a resolvent.
A resolvent $R(S)$ of $S$ is a chain of linear maps
\begin{equation}
0{\longrightarrow}\Gamma_{\!N}\stackrel{T_N}{\longrightarrow}\Gamma_{\!N-1}
\stackrel{T_{N-1}}{\longrightarrow}\dots\longrightarrow\Gamma_{\!1}
\stackrel{T_1}{\longrightarrow}ker(T){\longrightarrow}0 \label{1.4}
\end{equation}
with the property $T_kT_{k+1}=0$ for all $k=0,1,{\dots},N\,$. It determines
cohomology spaces
\begin{equation}
H^k(R(S))=ker(T_k)\,\Big/\,Im(T_{k+1}) \label{1.5}
\end{equation}
for all $k=0,1,\dots,N\,$.
The method given by Schwarz \cite{S} requires the cohomology of the resolvent
to vanish. In this case, given an inner product
$\langle\cdot\,,\cdot\rangle_k$ in each space $\Gamma_{\!k}\,$, the
volume $V(ker(T))$ in
(\ref{1.3}) can be formally evaluated from the resolvent
(\ref{1.4}) in terms of the divergent volumes $V(\Gamma_{\!k})\,$.
(This can be considered as a generalisation of the Faddeev-Popov method;
the details will be given in \S2).
An expression for the partition function is then obtained by substituting
the expression for $V(ker(T))$ into
(\ref{1.3}). It is normalised by dividing
out the divergent volumes $V(\Gamma_{\!k})\,$. The resulting expression is
\begin{equation}
Z(\beta)
=\pi^{\zeta/2}\,\beta^{-\zeta/2}\,det(\widetilde{T})^{-1/2}\,\prod_{k=1}^Ndet
(\widetilde{T}_k^*\widetilde{T}_k)^{\frac{1}{2}(-1)^{k-1}} \label{1.6}
\end{equation}
where $\widetilde{T}_k\,\colon\,ker(T_k)^{\perp}\stackrel{\simeq}
{\longrightarrow}Im(T_k)$ is the restriction of $T_k$ to $ker(T_k)^{\perp}\,$.
(This expression, without the factors $\pi^{\zeta/2}$ and $\beta^{-\zeta/2}$,
is the one obtained by Schwarz in \cite{S}).
When the resolvent is elliptic (defined in \S3) the determinants in
(\ref{1.6})
can be regularised by standard zeta-function regularisation
techniques as shown in \cite{S}.
As we will show, the quantity $\zeta$ in (\ref{1.6}) can also
be regularised by zeta-regularisation when the resolvent is elliptic.
Thus a finite expression for the partition function
(\ref{1.6}) is obtained.
It depends on the choice of resolvent (\ref{1.4}) and inner products
$\langle\cdot\,,\cdot\rangle_k$ in the $\Gamma_{\!k}\,$. We will show in
\S2 that in the finite-dimensional case (\ref{1.6})
has a symmetry-property under changes in the inner products
$\langle\cdot\,,\cdot\rangle_k$ which generalises a symmetry-property
of partition functions of non-degenerate functionals.

The main examples of quadratic functionals of interest from a topological
point of view (including those appearing in the semiclassical approximation
for the Witten-invariant) have canonical elliptic resolvents associated with
them. However, the cohomology of these resolvents is non-vanishing in
general. Hence the need to extend the method to the case where the
cohomology of the resolvent is non-vanishing (since no other methods exist
for dealing with this case). We do this in this paper. Our extension of
the method involves choosing inner products in the cohomology spaces
$H^k(R(S))$ and uses the maps
\begin{equation}
\Phi_k\,\colon\,{\cal H\/}_k\stackrel{\simeq}{\longrightarrow}H^k(R(S))
\label{1.7}
\end{equation}
obtained as the restriction of the projection maps $ker(T_k){\to}H^k(R(S))$
to ${\cal H}_k=Im(T_{k+1})^{\perp}{\cap}ker(T_k)\,$. (The details are
given in \S2). With a choice of orientation for each space $H^k(R(S))$
the dependence of the partition function on the inner products in the
$H^k(R(S))$ enters through the volume forms determined by the
orientations and inner products. Considered as a functional of these
volume forms the partition function can be interpreted as an element
\begin{equation}
Z(\beta)\,\in\,\otimes_{k=0}^N\Lambda^{max}H^k(R(S))^{*^{k+1}}\,. \label{1.8}
\end{equation}
(Here $W^*$ denotes the dual of a vectorspace $W$ and $W^{*^k}$ is identified
with $W$ or $W^*$ for $k$ even or odd respectively).

In the general case where $\beta\in{\bf C}$ and the quadratic functional $S$
is not required to be positive we formally evaluate the integral
(\ref{1.2}) in
\S2 and obtain a finite expression via analytic continuation in $\beta\,$.
The resulting expression for the partition function $Z(\beta)$ involves a
phase factor; this is well-defined for $\beta\in{\bf C}-{\bf R}\,$; for
$\beta\in{\bf R}$ there is an ambiguity analogous to the ambiguity in
$\sqrt{-1}={\pm}i\,$. (The case where $\beta=i\lambda$ is purely imaginary
is relevant for the semiclassical approximation for the Witten-invariant).
The following expression is obtained: For $\beta=|\beta|e^{i\theta}\in
{\bf C}_{\pm}$ with $\theta\in[-\pi,\pi]$ we get
\begin{eqnarray}
Z(\beta)
&=&\pi^{\zeta/2}\,e^{-\frac{i\pi}{4}((\frac{2\theta}{\pi}\mp1)\zeta\,\pm\,
\eta)}\,|\beta|^{-\zeta/2}\nonumber \\
& &\;\times\;det(\widetilde{T}^2)^{-1/4}\,\prod_{k=0}^N
\Bigl(det(\widetilde{T}_{k+1}^*\widetilde{T}_{k+1})det(\Phi_k^*\Phi_k)^{-1}
\Bigr)^{\frac{1}{2}(-1)^k}
\label{1.9}
\end{eqnarray}
In particular, for $\beta=i\lambda\;,\;\lambda>0\,$,
\begin{equation}
Z(\beta)
=\pi^{\zeta/2}\,e^{-\frac{i\pi}{4}\eta}\,\lambda^{-\zeta/2}\,det(\widetilde{T}
^2)^{-1/4}\,\prod_{k=0}^N\Bigl(det(\widetilde{T}_{k+1}^*\widetilde{T}_{k+1})
det(\Phi_k^*\Phi_k)^{-1}\Bigr)^{\frac{1}{2}(-1)^k}\,. \label{1.10}
\end{equation}
When the resolvent
(\ref{1.4}) is elliptic the determinants $det(\widetilde{T}^2)$
and $det(\widetilde{T}_k^*\widetilde{T}_k)$ are given well-defined finite
values via zeta-function regularisation, and the maps $\Phi_k$ (given
by (\ref{1.7})) are finite-dimensional.
The quantities $\zeta$ and $\eta$ are formally given by
$\zeta(0\,|\,|T|)$ and $\eta(0\,|\,T)\,$, where
$\zeta(s\,|\,|T|)$ and $\eta(s\,|\,T)$ are the zeta- and eta-functions
of $|T|$ and $T$ respectively (with $|T|=\sqrt{T^2}$ defined via
spectral theory).
These are given well-defined finite values via analytic continuation:
We show (theorems 3.2 and 3.3) that these
functions are regular at $s=0\,$, and that when $dimM$ is odd
\begin{equation}
\zeta=\zeta(0\,|\,|T|)=\sum_{k=0}^N(-1)^{k+1}dimH^k(R(S))\,. \label{1.11}
\end{equation}
Thus a well-defined expression for the partition function
(\ref{1.9}) is obtained
(up to a phase ambiguity for $\beta\in{\bf R}\,$). It depends on the choice
of elliptic resolvent $R(S)$ for $S$ and on choices of inner products in the
spaces $\Gamma_{\!k}$ and in the cohomology spaces $H^k(R(S))\,$.

For elliptic resolvents the spaces $\Gamma_{\!k}$ are the spaces of smooth
sections in vectorbundles over the manifold $M\,$, and the inner products
in the $\Gamma_{\!k}$ are constructed from Hermitian structures in the
bundles and a metric on $M\,$. In \cite{S} in the restricted case where
the cohomology of the resolvent vanishes and where $\beta$ is a constant
equal to 1 Schwarz derived formulae for the variation of (the modulus of)
the partition function under variation of the inner products in the
$\Gamma_{\!k}$ (theorem $1'$ in \cite{S}) and under a certain variation of
the maps $T_k$ in the resolvent (theorem $2'$ in \cite{S}). From these it
followed that when the (compact, closed) manifold $M$ has odd dimension
the partition function is invariant under these variations. In particular,
when the functional $S$ and resolvent $R(S)$ are topological (i.e. their
definitions do not require choices of Hermitian structures or metric on
$M\,$) the partition function is a topological invariant. We show in \S3
that these results continue to hold in the general case, i.e. for
non-vanishing cohomology and arbitrary $\beta\in{\bf C}\,$\footnote{The
metric-independence results are for the modulus of the partition function;
the phase is metric-dependent in general}. In addition we derive a new
invariance-property of the partition function under certain simultaneous
changes in the inner products and in the maps $T_k\,$.
The formula for the change in the partition function in this case is unlike
the previous ones in that its derivation involves only linear algebra as
opposed to elliptic operator theory (heat kernel expansion), and holds for
finite changes, not just infinitessimal changes. It relates the other
formulae (i.e. the generalisations of theorems $1'$ and $2'$ of Schwarz
\cite{S}) in that by combining it with the first
of these formulae we immediately obtain the second.

Finally we describe how the method enables the partition functions in the
semiclassical approximation for the Witten-invariant to be evaluated.
The Witten-invariant is formally given as the partition function of the
Chern-Simons gauge theory:
\begin{equation}
Z_W(k)=\int{\cal D\/}A\,e^{ikCS(A)}\;\;\;\;\;\;\;\;\;k\in{\bf Z} \label{1.12}
\end{equation}
where the Chern-Simons functional is
\begin{equation}
CS(A)=\frac{1}{4\pi}\int_M\mbox{Tr}(A{\wedge}dA+\frac{2}{3}A{\wedge}A
{\wedge}A) \label{1.13}
\end{equation}
The formal integration in
(\ref{1.12}) is over the gauge fields $A$ in a trivial
bundle, i.e. 1-forms on the 3-manifold $M$ with values in the
Lie algebra ${\bf g}$
of the gauge group $G\,$ (a compact Lie group). (The
expression
(\ref{1.13}) is for the case where $G=SU(N)\,$, identified with its
fundamental representation. For the correct expression in the general
case, and more information on the Chern-Simons functional, see \cite{F}.)
In the limit of large $k$ QFT predicts that (\ref{1.12}) is given by its
semiclassical approximation \cite[\S2]{W}:
\begin{equation}
\sum_{[A_f]}e^{ikCS(A_f)}\,{\int}{\cal D\/}\omega\,e^{\frac{ik}{4\pi}
\int_MTr(\omega{\wedge}d_{A_f}\omega)}\,. \label{1.14}
\end{equation}
The sum is over representatives $A_f$ for each point $[A_f]$ in the
moduli-space of flat gauge fields on $M\,$. The $\omega$ are
Lie-algebra-valued 1-forms and
$d_{A_f}$ is the covariant derivative determined by $A_f\,$, i.e.
$d_{A_f}\omega=d\omega+[A_f,\omega]\,$. In (\ref{1.14})
we are assuming that the moduli-space is
discrete, i.e. $H^1(d_{A_f})=0$ for all flat gauge fields $A_f\,$.
(When the moduli-space is not discrete then in certain cases
the sum in (\ref{1.14}) can be replaced by an
integral of a form of top degree on the
moduli-space, as pointed out by L.~Jeffrey \cite[\S5]{J}).
The summand in (\ref{1.14}) can be shown to be independent of the
choice of representative $A_f$ for each gauge-equivalence class.

The formal integrals in the semiclassical approximation
(\ref{1.14}) are partition
functions of quadratic functionals. They were partially evaluated by
Witten in \cite[\S2]{W} in the restricted case where the cohomology of
$d_{A_f}$ vanishes. Wittens method (which used gauge-fixing implemented via
a Lagrange-multiplier field) involves a ``rescaling'' of the fields
$\omega$ in
(\ref{1.14}) equivalent to setting $\frac{k}{4\pi}=1\,$, so the
$k-$dependence of the partition functions is not obtained\footnote{Actually
the method we give also leads to the partition function being independent
of $k$ when the cohomology of $d_{A_f}$ vanishes.}.

The method we describe in
this paper enables the partition functions in (\ref{1.14}) to be evaluated in
complete generality, i.e. the cohomology of $d_{A_f}$ is not required to
vanish and the dependence on the parameter $k$ is explicitly determined.
Each partition function in (\ref{1.14}) is of the form (\ref{1.1}) with
\begin{equation}
S(\omega)=\int_M-{\lambda_{\bf g}}Tr(\omega{\wedge}d_{A_f}\omega)
\;\;\;\;,\;\;\;\;\;\;
\beta=\frac{ik}{4\pi\lambda_{\bf g}} \label{1.15}
\end{equation}
where $\lambda_{\bf g}\in{\bf R}_+$
is an arbitrary parameter. There is a natural choice
of inner products in the spaces $\Omega^q(M,{\bf g})$ of ${\bf g}$-valued
q-forms in terms of which we have
$S(\omega)=\langle\omega\,,*d_{A_f(1)}\omega\rangle_{\lambda_{\bf g}}\,$.
The quadratic functional $S(\omega)$ in
(\ref{1.15}) has the canonical elliptic
resolvent
\begin{equation}
0\longrightarrow\Omega^0(M,{\bf g})\stackrel{d_{A_f(0)}}{\longrightarrow}
ker(d_{A_f(1)}){\longrightarrow}0 \label{1.16}
\end{equation}
The partition function is then given by (\ref{1.10}) with
$\lambda=\frac{k}{4\pi\lambda_{\bf g}}$ and $T=*d_{A_f(1)}\,$.
In particular, since the resolvent (\ref{1.16}) has cohomology spaces
$H^0(R(S))=H^1(d_{A_f})$ and $H^1(R(S))=H^0(d_{A_f})$ the $k-$dependence
of the partition function is given by (\ref{1.10}) and (\ref{1.11}) to be
\begin{equation}
\sim\,k^{-\frac{1}{2}(dimH^0(d_{A_f})-dimH^1(d_{A_f}))} \label{1.17}
\end{equation}
This is precisely the $k-$dependence conjectured by Freed and Gompf in
\cite{FG}, and as we show in \S4 the expression for the partition function
obtained from (\ref{1.10}) agrees with their conjectured expression up to an
overall numerical factor, except for Witten's geometric counterterms
in the phase.

The discrepancy in the numerical factor between the expression we obtain
and that conjectured in \cite{FG} is due in part to the factor
$\pi^{\zeta/2}$ in (\ref{1.3}), which was not taken into account in \cite{FG},
but this does not explain it completely. The
discrepancy can be understood from work by the first author \cite{Ad}
combined with results of Rozansky \cite{Roz}. We will discuss this in \S4
and illustrate it with an explicit calculation for $M=S^3\,$.

We mentioned previously that testing the prediction of QFT described
above is interesting from a mathematics point of view. The reason, as
pointed out in \cite{FG} \cite{FG1},
is as follows. The method and mathematical
machinery used to evaluate the Witten-invariant $Z_W(k)$ is very different
from that used to derive its semiclassical approximation. Evaluation
of the Witten-invariant is based on the axioms of topological QFT \cite{At}
\cite{At1}
and exploits a connection between Chern-Simons gauge theory in 3
dimensions and 2-dimensional conformal field theory. It draws on topology
(surgery techniques) and algebra (representation theory for Kac-Moody
algebras). On the other hand, deriving the semiclassical approximation
for the Witten-invariant by standard QFT methods, and evaluating it,
draws on differential geometry (gauge theory, Hodge theory) and analysis
(analytic continuation of functions). It is remarkable that through QFT
we obtain a link between these different areas of mathematics.

This paper is organised as follows:
In \S2 we study the partition function
(\ref{1.1}) in the case where $\Gamma$ is finite-dimensional (i.e. when the
dimension of the manifold is zero). This allows us to describe the method for
evaluating the partition function in a simple context. Formulae for the
change in the partition function under changes in the inner products and maps
in the resolvent are derived. We show that the method can be motivated by
requiring that a symmetry-property of the partition function of
non-degenerate quadratic functionals continues to hold in the degenerate case.
In \S3 the method is described for the case where $\Gamma$ is
infinite-dimensional and the resolvent is elliptic. Formulae for the
variation of the partition function under certain variations in the
structures used to construct it are derived (generalising results from \S2)
and resulting symmetry-properties are pointed out. These generalise results
of Schwarz \cite{S} and also lead to a new symmetry-property of the
partition function (theorem 3.10). Finally, in \S4 we apply the method
and results of \S3 to evaluate the partition functions of a particular class
of topological quadratic functionals. The resulting expression involves a
version of the Ray-Singer torsion as a ``function of the cohomology''
\cite[\S3]{RS2}. As a byproduct we show that the metric-dependence of the
usual Ray-Singer torsion factors out as a power of the volume of the
manifold in certain cases where the cohomology is non-vanishing (theorem 4.1).
The partition functions evaluated in \S4 include those appearing in the
semiclassical approximation for the Witten-invariant, and we show that the
expression obtained for these agrees with that conjectured by Freed and Gompf
\cite{FG} (modulo Witten's geometric counterterms in the phase, and a
discrepancy in the numerical factor which we explain). We explicitly
calculate the semiclassical approximation for the case of the 3-sphere
and show that it coincides with the Witten-invariant in the limit of
large $k\,$.

\section{The partition function in the finite-dimensional case}

\subsection{The partition function of a
positive quadratic functional}

We consider a quadratic functional $S(\omega)$ on a real
vectorspace\footnote{The method and results described in the following
also apply for real-valued quadratic functionals on complex vectorspaces,
see the remark at the end of this \S.}
of finite dimension $d\,$, set $\beta=1$
and assume that $S{\ge}0\,$. (We postpone the general case where
$\beta\in{\bf C}$ and $S$ is an arbitrary quadratic functional to the next
subsection).
An inner product
$\langle\cdot\,,\cdot\rangle_0$ in $\Gamma$ determines a measure
${\cal D\/}\omega$ on $\Gamma$ and allows us to uniquely
write $S(\omega)=
\langle\omega,T\omega\rangle_0$ with $T{\ge}0$ selfadjoint. If $S$ is
non-degenerate, i.e. if $S(\omega)>0$ for all $\omega{\ne}0$ then the
partition function (\ref{1.1}) of $S$ is well-defined and equals
\begin{equation}
Z(S,\langle\cdot\,,\cdot\rangle_0)=\pi^{d/2}\,det(T)^{-1/2} \label{2.1}
\end{equation}
This expression has the following symmetry-property. For each $A{\in}
GL(\Gamma)$ we obtain a new inner product $\langle\cdot\,,\cdot\rangle_0^A$
in $\Gamma$ from $\langle\cdot\,,\cdot\rangle_0$ by setting
\begin{equation}
{\langle}\upsilon,\omega\rangle_0^A={\langle}A\upsilon,A\omega\rangle_0
={\langle}\upsilon,A^*A\omega\rangle_0\,. \label{2.2}
\end{equation}
This determines a right group action of $GL(\Gamma)$ on the inner products
in $\Gamma\,$; note that any inner product can be obtained from an arbitrary
inner product $\langle\cdot\,,\cdot\rangle_0$ in this way. Under the action
of $A{\in}GL(\Gamma)$ the map $T$ is changed to $T_A\,$, where
\begin{equation}
S(\omega)={\langle}\omega,T_A\omega\rangle_0^A={\langle}\omega,A^*AT_A
\omega\rangle_0\;\;;\;\;\;T_A=(A^*A)^{-1}T\,. \label{2.3}
\end{equation}
Note that $T_A$ is selfadjoint w.r.t. $\langle\cdot\,,\cdot\rangle_0^A\,$.
It follows from (\ref{2.1}) that
\begin{equation}
Z(S,\langle\cdot\,,\cdot\rangle_0^A)=|det(A)|\,Z(S,\langle\cdot\,,
\cdot\rangle_0)\,. \label{2.4}
\end{equation}
Thus we see that for non-degenerate $S$ the partition function (\ref{2.1}) is
invariant under the group action of $SL|\Gamma|$ on $\langle\cdot\,,
\cdot\rangle_0\,$, where $SL|\Gamma|$ denotes the subgroup of $GL(\Gamma)$
consisting of the $A$ with $|det(A)|=1\,$.

If the functional $S$ is degenerate then $ker(T){\ne}0$ and the partition
function is given by the expression (\ref{1.3}), which diverges due to the
divergent volume factor $V(ker(T))\,$. To obtain a finite expression in this
case the partition function must be normalised. The simplest way to do this
is to divide out (i.e. discard) the factor $V(ker(T))$ in (\ref{1.3});
this gives $Z=\pi^{\zeta/2}\,det(\widetilde{T})^{-1/2}\,$.
However, this expression does not have the invariance-property under the
action of $SL|\Gamma|$ on $\langle\cdot\,,\cdot\rangle_0$ described above.
We can preserve this invariance-property by proceeding as follows.
Choose a resolvent $R(S)$ for $S$ as in (\ref{1.4}), i.e. a chain of linear
maps between real finite-dimensional vectorspaces:
\begin{equation}
0\longrightarrow\Gamma_{\!N}\stackrel{T_N}{\longrightarrow}\Gamma_{\!N-1}
\stackrel{T_{N-1}}{\longrightarrow}\dots\longrightarrow\Gamma_1
\stackrel{T_1}{\longrightarrow}ker(S){\longrightarrow}0 \label{2.5}
\end{equation}
with the property that $T_kT_{k+1}=0$ for all $k=0,1,\dots,N\,$. (Note
that $ker(S)=S^{-1}(0)$ is the same as $ker(T)$ since we are assuming
$S{\ge}0\,$.) We assume to begin with that the cohomology spaces (\ref{1.5})
of the resolvent all vanish. Then, choosing an inner product $\langle\cdot\,,
\cdot\rangle_k$ in each $\Gamma_{\!k}\,$, the resolvent
(\ref{2.5}) enables us to
formally calculate $V(ker(S))$ in terms of the divergent volumes
$V(\Gamma_{\!k})\,$: The maps $T_k$ restrict to maps $\widetilde{T}_k
\,\colon\,ker(T_k)^{\perp}\stackrel{\simeq}{\longrightarrow}
Im(T_k)\,$, which leads to the
formal relation
\begin{equation}
V(Im(T_k))=|det(\widetilde{T}_k)|\,V(ker(T_k)^{\perp})=det(\widetilde{T}_k^*
\widetilde{T}_k)^{1/2}\,V(ker(T_k)^{\perp})\,. \label{2.6}
\end{equation}
It is possible that $\widetilde{T}_k=0\,$ and we define the determinant of the
zero-map on the zero-dimensional vectorspace to be equal to 1 here and
in the following.
{}From the orthogonal decomposition $\Gamma_{\!k}=ker(T_k){\oplus}ker(T_k)^
{\perp}$ we get the formal relation
\begin{equation}
V(\Gamma_{\!k})=V(ker(T_k)){\cdot}V(ker(T_k)^{\perp})\,. \label{2.7}
\end{equation}
Combining this with (\ref{2.6}) gives
\begin{equation}
V(Im(T_k))=det(\widetilde{T}_k^*\widetilde{T}_k)^{1/2}\,V(\Gamma_{\!k})
\,V(ker(T_k))^{-1}\,. \label{2.8}
\end{equation}
The assumption of vanishing cohomology means $ker(T_k)=Im(T_{k+1})\,$,
so from (\ref{2.8})
\begin{equation}
V(ker(T_k))=det(\widetilde{T}_{k+1}^*\widetilde{T}_{k+1})^{1/2}\,
V(\Gamma_{\!k+1})\,V(ker(T_{k+1}))^{-1}\,. \label{2.9}
\end{equation}
Now a simple induction argument based on (\ref{2.9}) gives
\begin{equation}
V(ker(S))=\prod_{k=1}^N\,
\Bigl(det(\widetilde{T}_k^*\widetilde{T}_k)^{1/2}
\,V(\Gamma_{\!k})\Bigr)^{(-1)^{k-1}}\,. \label{2.10}
\end{equation}
Substituting this for $V(ker(T))$ in (\ref{1.3}) gives a formal expression for
the partition function. We normalise this expression by dividing out the
divergent volumes $V(\Gamma_{\!k})\,$. This gives
\begin{equation}
Z(R(S),\langle\cdot\,,\cdot\rangle)=
\pi^{\zeta/2}\,det(\widetilde{T})^{-1/2}\,\prod_{k=1}^Ndet(\widetilde{T}_k^*
\widetilde{T}_k)^{\frac{1}{2}(-1)^{k-1}}\,. \label{2.11}
\end{equation}
This procedure for evaluating and normalising
the partition function (\ref{1.1}) to obtain (\ref{2.11}) is essentially
(the finite-dimensional version of) Schwarz's method \cite{S}.
The partition
function (\ref{2.11}) depends on the choice $R(S)$ of resolvent
(\ref{2.5}) and
choice of inner products $\langle\cdot\,,\cdot\rangle_k$ in the spaces
$\Gamma_{\!k}\,$, which we collectively denote by $\langle\cdot\,,\cdot
\rangle\,$.

The group action (\ref{2.2}) of $GL(\Gamma)$ on $\langle\cdot\,,\cdot\rangle_0$
generalises to a right group action of $GL(\Gamma_{\!N})\times\cdots{\times}
GL(\Gamma_{\!1}){\times}GL(\Gamma)$ on $\langle\cdot\,,\cdot\rangle\,$:
An element $A\!=\!(A_N,\dots,A_0)$ determines inner products $\langle\cdot\,,
\cdot\rangle^A=\lbrace\langle\cdot\,,\cdot\rangle_N^{A_N},\dots,\langle
\cdot\,,\cdot\rangle_0^{A_0}\rbrace$ with $\langle\cdot\,,\cdot
\rangle_k^{A_k}$ defined by analogy with
(\ref{2.2}). The following theorem is a
generalisation of (\ref{2.4}) above; it shows that the invariance-property of
the partition function of non-degenerate $S$ generalises for the partition
function obtained from the above method in the degenerate case.

\vspace {1ex}

\noindent {\bf Theorem 2.1.} {\it
Let $A=(A_N,\dots,A_1,A_0){\in}\,GL(\Gamma_N){\times}\dots{\times}
GL(\Gamma_1){\times}GL(\Gamma)$. The action
of $A$ on the inner products $\langle{\cdot}\,,
\cdot\rangle$ changes the partition function (\ref{2.11}) to}
\begin{equation}
Z(R(S),\langle\cdot\,,\cdot\rangle^A)=
\biggl(\,\prod_{k=0}^N|det(A_k)|^{(-1)^k}\biggr)\,
Z(R(S),\langle\cdot\,,\cdot\rangle)
\label{2.12}
\end{equation}

\vspace {1ex}
\noindent {\bf Corollary 2.2.} {\it
The partition function
(\ref{2.11}) is invariant
under the action of $SL|\Gamma_{\!N}|\times\dots{\times}
SL|\Gamma_{\!1}|{\times}SL|\Gamma|$ on the inner products $\langle{\cdot}\,,
\cdot\rangle$.}

\vspace {1ex}

The proof of the theorem rests on a celebrated theorem of Jacobi from 1834.
We introduce the following notation:
If $L\,\colon\,V{\longrightarrow}V$ is a linear map and $P$ is a
projection map on $V$ then by $det(PLP)$ we mean the determinant of the
restriction of $PLP$ to the
map $PLP\,\colon\,Im(P){\longrightarrow}Im(P)$.
For each map $T_k\,\colon\,\Gamma_{\!k}\longrightarrow\Gamma_{\!k-1}$ in the
resolvent
(\ref{2.5}) define $P_k$ and $Q_k$ to be the orthogonal projections of
$\Gamma_{\!k}$ on $ker(T_k)$ and $ker(T_k)^{\perp}$ respectively.
Let $T_k^{*(A)}$ denote the adjoint of $T_k$ determined by the
inner products $\langle\cdot\,,\cdot\rangle_k^{A_k}$ and $\langle\cdot\,,\cdot
\rangle_{k-1}^{A_{k-1}}\,$. For subspace $W\subseteq\Gamma_{\!k}$ let
$W^{\perp(A)}$ denote the orthogonal complement of $W$ determined by
$\langle\cdot\,,\cdot\rangle_k^{A_k}\,$.

The action of $A$ on the inner products changes $det(\widetilde{T})$ to
$det(\widetilde{T}_{A_0})\,$, where $\widetilde{T}_{A_0}$ is the
restriction of $T_{A_0}\,$ (defined as in (\ref{2.3})) to $ker(T_{A_0})^
{\perp(A)}\,$, and changes $det(\widetilde{T}_k^*\widetilde{T}_k)$ to
$det(\widetilde{T}_{k(A)}^{*(A)}\widetilde{T}_{k(A)})\,$, where
$\widetilde{T}_{k(A)}$ is the restriction of $T_k$ to
$ker(T_k)^{\perp(A)}\,$. In the appendix we show that
\begin{equation}
det(\widetilde{T}_{A_0})
=det(Q_0(A_0^*A_0)^{-1}Q_0)\,det(\widetilde{T}) \label{2.13}
\end{equation}
and
\begin{equation}
det(\widetilde{T}_{k(A)}^{*(A)}\widetilde{T}_{k(A)})=
\,det(Q_k(A_k^*A_k)^{-1}Q_k)\,
det(P_{k-1}(A_{k-1}^*A_{k-1})P_{k-1})\,
det(\widetilde{T}_k^*\widetilde{T}_k). \label{2.14}
\end{equation}
It follows that the partition function (\ref{2.11})
is changed to
\begin{eqnarray}
\lefteqn{Z(R(S),\langle\cdot\,,\cdot\rangle^A)} \nonumber\\
&=&\prod_{k=0}^N\Bigl(\,det(Q_k(A_k^*A_k)^{-1}Q_k)^{-1}\,
det(P_k(A_k^*A_k)P_k)\Bigr)^{\frac{1}{2}(-1)^k}
\,Z(R(S),\langle\cdot\,,\cdot\rangle)\,. \nonumber \\
& & \;\label{2.15}
\end{eqnarray}
We can obtain (\ref{2.12}) from (\ref{2.15}) using the following

\vspace {1ex}

\noindent {\bf Lemma 2.3.} {\it
 Let $L\,\colon\,V{\longrightarrow}V$ be an invertible linear map on a finite-
\break dimensional vectorspace,
and let $P\,\colon\,V{\longrightarrow}V$ be a projection. Set $Q=I-P$,
where $I$ is the identity map. Then we have the formula}
\begin{equation}
det(PLP)=det(QL^{-1}Q)\,det(L)\,.
\end{equation}

\vspace{1ex}

\noindent This is presumably a classical result in the study of determinants;
it is an easy consequence of the aforementioned
theorem of Jacobi. This theorem can be found in
e.g. \cite[pp98--99]{Ai}.
Note that if $L$ is strictly positive and selfadjoint with respect to some
inner product in $V$ then $PLP$ and $QL^{-1}Q$ are also strictly positive
and selfadjoint, and therefore invertible. It follows that in this case we
have
\begin{equation}
det(PLP)\,det(QL^{-1}Q)^{-1}=det(L)\,. \label{2.16}
\end{equation}
Using (\ref{2.16}) with $L=A_k^*A_k$ and $P=P_k$ we see
that (\ref{2.15}) is equal to (\ref{2.12}), proving the theorem.

In order to compare with the infinite-dimensional case we give
the expression for the variation of the
partition function (\ref{2.11}) under
variation of the inner products. Let $A(t)=(A_N(t),{\dots},A_0(t))$ be a
smooth curve in $GL(\Gamma_{\!N}){\times}
\cdots{\times}GL(\Gamma)$ with $A(0)=I$ (the identity),
and set $B_k=\;$ \break
$\frac{d}{dt}\Bigl|_{t=0}
(A_k^*(t)A_k(t))$ for $k=0,1,{\dots},N$,
then $\frac{d}{dt}\Bigl|_{t=0}{\langle}u,v\rangle_k^{A_k(t)}=
{\langle}u,B_kv\rangle_k$. For $t{\rightarrow}0$ we have
$$
(A_k^*(t)A_k(t))^{1/2}=1+\frac{1}{2}tB_k+O(t^2)=e^{\frac{1}{2}tB_k}+O(t^2)
$$
which leads to
$$
\frac{d}{dt}\biggl|_{t=0}|det(A_k(t))|=
\frac{d}{dt}\biggl|_{t=0}det(A_k^*(t)A_k(t))^{1/2}
=\frac{1}{2}\mbox{Tr}(B_k)\,.
$$
It now follows from (\ref{2.12}) that
\begin{equation}
\frac{d}{dt}\biggl|_{t=0}Z(R(S),\langle\cdot\,,\cdot\rangle^{A(t)})
=\biggl(\sum_{k=0}^N\frac{1}{2}(-1)^kTr(B_k)\biggr)\,
Z(R(S),\langle\cdot\,,\cdot\rangle)\,.
\label{2.18}
\end{equation}

We now consider the case where the cohomology (\ref{1.5})
of the resolvent (\ref{2.5}) is non-vanishing. In this case
the expression (\ref{2.11}) for the partition
function is no longer formally correct because  we no longer have
$V(ker(T_k))=V(Im(T_{k+1}))$ in the formal
calculation leading to (\ref{2.9}). Note that if one simply defines the
partition function by (\ref{2.11}) in this case
then the invariance of the partition function under the action
of $SL|\Gamma_{\!N}|{\times}\cdots{\times}
SL|\Gamma_{\!1}|{\times}SL|\Gamma|$ on the inner products
no longer holds. We now describe how to construct
the partition function in a way which maintains its invariance-properties
when the cohomology of the resolvent is non-vanishing.
In fact we will show that in this case the partition
function can be interpreted as an element in
$\otimes_{k=0}^N\Lambda^{max}H^k(R(S))^{*^{k+1}}\,$.
(Here and in the following $W^*$ is the dual
of a vectorspace $W\,$, and $W^{*^k}$ is
identified with $W$ or $W^*$ for $k$ even or odd
respectively).

Given a resolvent (\ref{2.5})
and inner products $\langle\cdot\,,\cdot\rangle_k$
in the $\Gamma_{\!k}$
we define  ${\cal H\/}_k{\subseteq}ker(T_k)$
by the following orthogonal decomposition:
\begin{equation}
{\ker}(T_k)=Im(T_{k+1}){\oplus}{\cal H\/}_k \,. \label{2.19}
\end{equation}
The projection map $ker(T_k){\longrightarrow}
H^k(R(S))$ restricts to an isomorphism
\begin{equation}
\Phi_k\,\colon\,{\cal H\/}_k\stackrel{\simeq}{\longrightarrow}H^k(R(S))\,.
\label{2.20}
\end{equation}
The calculations leading to (\ref{2.10})
can now be modified to formally evaluate the
volume $V(ker(S))$ in the case where the
cohomology of the resolvent is non-vanishing.
First, from (\ref{2.19}) we get the formal relation
\begin{equation}
V(ker(T_k))=V(Im(T_{k+1})){\cdot}V({\cal H\/}_k)\,. \label{2.21}
\end{equation}
Next we pick an inner product $\langle\cdot\,,\cdot\rangle_{H^k}$
in each of the spaces $H^k(R(S))$ and get from (\ref{2.20})
the formal relation
\begin{equation}
V({\cal H\/}_k)=|det(\Phi_k^{-1})|\,V(H^k(R(S)))
=det(\Phi_k^*\Phi_k)^{-1/2}\,V(H^k(R(S))) \,.\label{2.22}
\end{equation}
Substituting this into (\ref{2.21}) a straightforward modification of the
calculations leading to (\ref{2.10}) gives
\begin{equation}
V(ker(S))=\prod_{k=0}^N\biggl(\!det(\widetilde{T}_{k+1}^*
\widetilde{T}_{k+1})^{1/2}det(\Phi_k^*\Phi_k)
^{-1/2}\,V(\Gamma_{\!k+1})V(H^k(R(S)))\!\biggr)^{(-1)^k} \label{2.23}
\end{equation}
Substituting this for $V(ker(T))$ in (\ref{1.3})
gives a formal expression for the partition function. We normalise this
expression by dividing out the divergent
volumes $V(\Gamma_{\!k})$ and $V(H^k(R(S)))\,$. This gives
\begin{eqnarray}
\lefteqn{
Z(R(S),\langle\cdot\,,\cdot\rangle_H,\langle\cdot\,,
\cdot\rangle)}\nonumber \\
&=&\pi^{\zeta/2}det(\widetilde{T})^{-1/2}
\prod_{k=0}^N\biggl(det(\widetilde{T}_{k+1}^*\widetilde{T}_{k+1})
^{1/2}\,
det(\Phi_k^*\Phi_k)^{-1/2}\biggr)^{(-1)^k}\,.\label{2.24}
\end{eqnarray}
This expression
depends not only on the resolvent $R(S)$ and the inner products
$\langle\cdot\,,\cdot\rangle$
in the $\Gamma_{\!k}$,
but also on the inner products $\langle\cdot\,,
\cdot\rangle_{H^k}\,$ in the $H^k(R(S))\,$, which we collectively
denote by $\langle\cdot\,,\cdot\rangle_H\,$.
The expression (\ref{2.24}) is proportional to
\begin{equation}
\Psi(R(S),\langle\cdot\,,\cdot\rangle_H,\langle\cdot\,,\cdot\rangle)
=\prod_{k=0}^Ndet(\Phi_k^*\Phi_k)^{\frac{1}{2}(-1)^{k+1}}\,. \label{2.25}
\end{equation}
Considered as a functional of the inner products
$\langle\cdot\,,\cdot\rangle_{H^k}$
in the $H^k(R(S))$ we can interpret $\Psi$ as an element $\widehat{\Psi}$
in $\otimes_{k=0}^N\Lambda^{max}H^k(R(S))^{*^{k+1}}\,$:
The map $\Phi_k$ in (\ref{2.20}) induces maps
\begin{eqnarray*}
\hat{\Phi}_k&:&\Lambda^{max}{\cal H\/}_k\stackrel{\simeq}{\longrightarrow}
\Lambda^{max}
H^k(R(S)) \\
(\hat{\Phi}_k^*)^{-1}&:&\Lambda^{max}{\cal H\/}_k^*
\stackrel{\simeq}{\longrightarrow}
\Lambda^{max}H^k(R(S))^*\,.
\end{eqnarray*}
Now fix orientations for the cohomology spaces
$H^k(R(S))\,$, these induce orientations for
the spaces ${\cal H\/}_k$ via the maps
$\Phi_k$ which together with the inner products
$\langle\cdot\,,\cdot\rangle_k$ determine volume
elements $w_k\in\Lambda^{max}{\cal H\/}_k$
and $w_k^*\in\Lambda^{max}{\cal H\/}_k^*\,$. We define
\begin{equation}
\widehat{\Psi}(R(S),\langle\cdot\,,\cdot\rangle)=
\otimes_{k=0}^N
(\widehat{\Phi}_k^{*^{k+1}})^{(-1)^{k+1}}(w_k^{*^{k+1}})\,{\in}\,
\otimes_{k=0}^N\Lambda^{max}H^k(R(S))^{*^{k+1}}\,. \label{2.26}
\end{equation}
The inner product $\langle\cdot\,,\cdot\rangle_{H^k}$ and orientation
in each $H^k(R(S))$ determine volume
elements $v_k\in\Lambda^{max}H^k(R(S))$ and
$v_k^*\in\Lambda^{max}H^k(R(S))^*\,$. Define
\begin{equation}
v_H=\otimes_{k=0}^Nv_k^{*^k}\,\in\,
\otimes_{k=0}^N\Lambda^{max}H^k(R(S))^{*^k}\,. \label{2.27}
\end{equation}
Let $<\!\cdot\,,\!\cdot>_k$ denote the natural
pairing of $\Lambda^{\!max}H^k(\!R(S))$ and $\Lambda^{\!max}H^k(\!R(S))^*\,$
then
$$
<\widehat{\Phi}_k(w_k),v_k^*>_k=det(\Phi_k^*\Phi_k)^{1/2}\;\;,\;\;
<(\widehat{\Phi}_k^*)^{-1}(w_k^*),v_k>_k=det(\Phi_k^*\Phi_k)^{-1/2}
$$
and it follows, with $<\cdot\,,\cdot>=\otimes_{k=0}^N<\cdot\,,
\cdot>_k\,$, that
$$
\Psi(R(S),\langle\cdot\,,\cdot\rangle_H,\langle\cdot\,,\cdot\rangle)=
<\widehat{\Psi}(R(S),\langle\cdot\,,\cdot\rangle),v_H>\,.
$$
Thus we see that $\Psi$ in (\ref{2.25}) can be identified with
$\widehat{\Psi}$ in (\ref{2.26}).
It follows that as a functional of the inner products $\langle\cdot\,,
\cdot\rangle_H$ in the cohomology spaces $H^*(R(S))$ the partition function
can be considered as an element
\begin{equation}
\widehat{Z}(R(S),\langle\cdot\,,\cdot\rangle)\,\in\,\otimes_{k=0}^N
\Lambda^{max}H^k(R(S))^{*^{k+1}} \label{2.28}
\end{equation}
with
\begin{equation}
Z(R(S),\langle\cdot\,,\cdot\rangle_H,\langle\cdot\,,\cdot\rangle)=
<\widehat{Z}(R(S),\langle\cdot\,,\cdot\rangle),v_H>\,. \label{2.29}
\end{equation}

The partition function (\ref{2.24}) has the
same invariance-property as before under change of the inner
products in the spaces $\Gamma_{\!k}$, as the following theorem shows.

\vspace{1ex}

\noindent {\bf Theorem 2.4.} {\it
Let $A=(A_N,\dots,A_1,A_0)\in\,GL(\Gamma_{\!N})
{\times}\cdots{\times}GL(\Gamma_{\!1}){\times}
GL(\Gamma)$. The action of $A$ on the
inner products in the spaces $\Gamma_{\!k}$ changes the
partition function (\ref{2.24}) to}
\begin{equation}
Z(R(S),\langle\cdot\,,\cdot\rangle_H,\langle\cdot\,,\cdot\rangle^A)
=\biggl(\,\prod_{k=0}^N|det(A)|^{(-1)^k}\biggr)\,
Z(R(S),\langle\cdot\,,\cdot\rangle_H,\langle\cdot\,,\cdot\rangle) \label{2.30}
\end{equation}

\vspace {1ex}
\noindent {\bf Corollary 2.5.} {\it
The partition function (\ref{2.24}) is invariant
under the action of $SL|\Gamma_{\!N}|{\times}\cdots{\times}
SL|\Gamma_{\!1}|{\times}SL|\Gamma|$ on the
inner products in the spaces $\Gamma_{\!k}$.}

\vspace {1ex}

\noindent Theorem 2.4 and
corollary 2.5 also hold when the partition function (\ref{2.24})
is replaced by (\ref{2.28}).

To prove the theorem we show in the appendix that the action of $A$ on
$\langle\cdot\,,\cdot\rangle$ changes $det(\widetilde{T}_k^*\widetilde{T}_k)\,
det(\Phi_{k-1}^*\Phi_{k-1})^{-1}$ to
\begin{eqnarray}
\lefteqn{det(\widetilde{T}_{k(A)}^{*(A)}\widetilde{T}_{k(A)})\,
det(\Phi_{k-1(A)}^{*(A)}\Phi_{k-1(A)})^{-1}}\nonumber \\
&=&det\Bigl(Q_k(A_k^*A_k)^{-1}Q_k\Bigr)\,
det\Bigl(P_{k-1}(A_{k-1}^*A_{k-1})P_{k-1}
\Bigr)\nonumber \\
& &\;\times\;det(\widetilde{T}_k^*\widetilde{T}_k)\,det(\Phi_{k-1}^*\Phi_{k-1})
^{-1}\,.\label{2.31}
\end{eqnarray}
The theorem now follows from
(\ref{2.13}) and (\ref{2.31}) in the same way that theorem 2.1
followed from (\ref{2.13}) and (\ref{2.14}).

The variation of the partition function
(\ref{2.24}) under variation of the inner products in the spaces
$\Gamma_{\!k}$ can be derived from
(\ref{2.30}) in the same way that (\ref{2.18})
was derived from (\ref{2.12}). The
result is
\begin{equation}
\frac{d}{dt}\biggl|_{t=0}Z(R(S),\langle\cdot\,,\cdot\rangle_H,
\langle\cdot\,,\cdot\rangle^{A(t)})=
\biggl(\sum_{k=0}^N\frac{1}{2}(-1)^kTr(B_k)\biggr)
Z(R(S),\langle\cdot\,,\cdot\rangle_H,\langle\cdot\,,\cdot\rangle)
\label{2.32}
\end{equation}
where $A(t)$ and $B_k$ are as in (\ref{2.18}).

In \cite{S}
the dependence of the partition function on the maps $T_k$ in the resolvent
was investigated, which we now discuss.
Given a resolvent $R(S)$ as in (\ref{2.5}) we can construct new
resolvents as follows.
For any $C=(C_N,\dots,C_1,\;$ \break
$C_0)\in\,
GL(\Gamma_{\!N}){\times}\cdots{\times}GL(\Gamma_{\!1}){\times}
GL(ker(S))$
we define maps $T_k^C=\;$ \break
$C_{k-1}^{-1}{\circ}T_k{\circ}C_k\,\colon\,\Gamma_{\!k}
\longrightarrow\Gamma_{\!k-1}$
for each $k=1,\dots,N\,$. Then by replacing each $T_k$
in (\ref{2.5}) by $T_k^C$ we obtain a new resolvent
for the functional $S$, which we denote by $R(S)^C$.
This defines a right-action of $GL(\Gamma_{\!N}){\times}
\cdots{\times}GL(\Gamma_{\!1}){\times}GL(ker(S))$ on the set of
resolvents for the functional $S$.
Note that $ker(T_k^C)=\;$ \break
$C_k^{-1}(ker(T_k))$ and
$Im(T_{k+1}^C)=
C_k^{-1}(Im(T_{k+1}))\,$, it follows that
$C_k^{-1}$ on \break
$ker(T_k)$ induces an isomorphism
\begin{equation}
C_k^{-1}\,\colon\,H^k(R(S)^C)\stackrel{\simeq}
{\longrightarrow}H^k(R(S))\,. \label{2.33}
\end{equation}
It follows that the dimensions of the cohomology spaces of $R(S)^C$ are the
same as for $R(S)\,$.
Conversely, if $R(S)'$ is another resolvent with the same
spaces $\Gamma_{\!k}$ and same dimensions of the cohomology spaces as
$R(S)\,$ then it is easy to see that $R(S)'=R(S)^C$ for some $C\,$.
Given a collection $\langle\cdot\,,\cdot\rangle_H=\lbrace\langle\cdot\,,\cdot
\rangle_{H^0},\dots,
\langle\cdot\,,\cdot\rangle_{H^N}\rbrace$ of inner products
in the cohomology spaces $H(R(S))$ it follows from (\ref{2.33}) that
$\langle\cdot\,,\cdot\rangle_{H^k}^{C_k}\,$, defined by analogy with
(\ref{2.2}), defines an inner product in $H^k(R(S)^C)\,$;
we denote the collection
of these by $\langle\cdot\,,\cdot\rangle_H^C\,$.
The group $GL(\Gamma_{\!N})\times\dots{\times}GL(\Gamma_{\!1}){\times}
GL(ker(S))$ also acts on the inner products $\langle\cdot\,,\cdot\rangle$
in the spaces $\Gamma_{\!k}\,$. (We extend $C_0{\in}GL(ker(S))$ to
$GL(\Gamma)$ by defining $C_0$ to be the identity map on $ker(S)^{\perp}\,$).
The partition function
has the following invariance-property:

\vspace{1ex}

\noindent {\bf Theorem 2.6.} {\it
For all $C=(C_N,\dots,C_1,C_0)\;{\in}\;GL(\Gamma_{\!N})\times\cdots{\times}
GL(\Gamma_{\!1}){\times}\;$ \break
$GL(ker(S))$
the partition function (\ref{2.24}) satisfies}
\begin{equation}
Z(R(S)^C,\langle\cdot\,,\cdot\rangle_H^C,\langle\cdot\,,\cdot\rangle^C)=
Z(R(S),\langle\cdot\,,\cdot\rangle_H,\langle\cdot\,,\cdot\rangle)\,.
\label{2.34}
\end{equation}

\vspace{1ex}

\noindent {\it Proof.} Let $(T_k^C)^{*(C)}$ denote the adjoint of $T_k^C$
w.r.t. the inner products $\langle\cdot\,,\cdot\rangle_k^{C_k}$ and
$\langle\cdot\,,\cdot\rangle_{k-1}^{C_{k-1}}\,$, and for a subspace
$W{\subseteq}\Gamma_{\!k}$ let $W^{\perp(C)}$ denote the orthogonal
complement of $W$ w.r.t. $\langle\cdot\,,\cdot\rangle_k^{C_k}\,$.
For $v\in\Gamma_{\!k-1}\,,\,w\in\Gamma_{\!k}\,$,
$$
{\langle}v,T_k^Cw\rangle_{k-1}^{C_{k-1}}=
{\langle}C_k^{-1}T_k^*C_{k-1}v,w\rangle_k^{C_k}
$$
so
\begin{equation}
(T_k^C)^{*(C)}T_k^C=C_k^{-1}T_k^*T_kC_k\,. \label{2.35}
\end{equation}
Let $\widetilde{T}_k^C$ denote the restriction of $T_k^C$ to
$ker(T_k^C)^{\perp(C)}=C_k^{-1}(ker(T_k)^{\perp})\,$, then from (\ref{2.35})
\begin{equation}
det((\widetilde{T}_k^C)^{*(C)}\widetilde{T}_k^C)=
det(\widetilde{T}_k^*\widetilde{T}_k)\,. \label{2.36}
\end{equation}
Since $C_0$ is the identity map on $ker(S)^{\perp}$ we see from
(\ref{2.3}) that
$\widetilde{T}_{C_0}=\widetilde{T}\,$, so
\begin{equation}
det(\widetilde{T}_{C_0})=det(\widetilde{T})\,. \label{2.36a}
\end{equation}
Let $\Phi_k^C\,\colon\,{\cal H\/}_k^C\stackrel{\simeq}{\longrightarrow}
H^k(R(S)^C)$ denote the map constructed by analogy with (\ref{2.20}) from
$\langle\cdot\,,\cdot\rangle^C$ and $R(S)^C\,$, then ${\cal H\/}_k^C
=C_k^{-1}({\cal H\/}_k)\,$, $H^k(R(S)^C)=$ \break
$C_k^{-1}(H^k(R(S))$ and
$\Phi_k^C=C_k^{-1}\Phi_kC_k\,$. By an argument similar to that leading
to (\ref{2.36}) we obtain
\begin{equation}
(\Phi_k^C)^{*(C)}\Phi_k^C=C_k^{-1}\Phi_k^*\Phi_kC_k \label{2.37}
\end{equation}
and therefore
\begin{equation}
det((\Phi_k^C)^{*(C)}\Phi_k^C)=det(\Phi_k^*\Phi_k)\,. \label{2.38}
\end{equation}
Combining (\ref{2.36}), (\ref{2.36a})
and (\ref{2.38}) with the expression (\ref{2.24}) gives (\ref{2.34}),
proving the theorem.

Combining theorem 2.6 with theorem 2.4 allows to determine the change
in the partition function resulting from the group action on the pair
$\;(R(S),\;$ \break
$\langle\cdot\,,\cdot\rangle_H)\,$: From theorem 2.6,
$$
Z(R(S)^C,\langle\cdot\,,\cdot\rangle_H^C,\langle\cdot\,,\cdot\rangle)=
Z(R(S),\langle\cdot\,,\cdot\rangle_H,\langle\cdot\,,\cdot\rangle^{C^{-1}})
$$
and theorem 2.4 now gives

\vspace{1ex}

\noindent {\bf Corollary 2.7}
\begin{equation}
Z(R(S)^C,\langle\cdot\,,\cdot\rangle_H^C,\langle\cdot\,,\cdot\rangle)=
\biggl(\prod_{k=0}^N|det(C_k)|^{(-1)^{k+1}}\biggr)\,
Z(R(S),\langle\cdot\,,\cdot\rangle_H,\langle\cdot\,,\cdot\rangle)\,.
\label{2.39}
\end{equation}
{\it
In particular $Z(R(S),\langle\cdot\,,\cdot\rangle_H,\langle\cdot\,,
\cdot\rangle)$ is invariant under the action of $SL|\Gamma_{\!N}|
\times\cdots{\times}SL|\Gamma_{\!1}|{\times}SL|ker(S)|$ on the
pair
$\;(R(S),\langle\cdot\,,\cdot\rangle_H)\,$.}

\vspace{1ex}

This allows to derive the variation of the partition function under variation
of the pair $(R(S),\langle\cdot\,,\cdot\rangle_H)$ by the group action:
Let $C(t)$ be a smooth curve in $GL(\Gamma_{\!N})\times\cdots{\times}
GL(\Gamma_{\!1}){\times}GL(ker(S))$
with $C(0)\!=\!I$ and set $D_k\!=\!\frac{d}{dt}\biggl|_{t=0}
(C_k^*(t)C_k(t))$. Then from (\ref{2.39}),
\begin{eqnarray}
\lefteqn{\frac{d}{dt}\biggl|_{t=0}Z(R(S)^{C(t)},\langle\cdot\,,
\cdot\rangle_H^{C(t)},\langle\cdot\,,\cdot\rangle)}\nonumber \\
&=&\biggl(\sum_{k=0}^N\frac{1}{2}(-1)^{k+1}\mbox{Tr}(D_k)\biggr)\,
Z(R(S),\langle\cdot\,,\cdot\rangle_H,\langle\cdot\,,\cdot\rangle)\,.
\label{2.40}
\end{eqnarray}

The results derived in this subsection may be of independent algebraic
interest. The partition function (\ref{2.24}) is the square root of the norm of
the determinant of the following complex of linear maps\footnote{ For a
description of the determinant of a complex of linear maps see
e.g. \cite[appendix A]{Gelf}.} :
$$
0\longrightarrow\Gamma_{\!N}\stackrel{T_N}{\longrightarrow}\dots
\longrightarrow\Gamma_{\!1}
\stackrel{T_1}{\longrightarrow}\Gamma\stackrel{T}{\longrightarrow}\Gamma
\stackrel{T_1^*}{\longrightarrow}
\Gamma_{\!1}\longrightarrow\dots\stackrel{T_N^*}{\longrightarrow}
\Gamma_{\!N}{\longrightarrow}0\,.
$$
In the special case where $S=0$ the partition function (\ref{2.28}) is the
determinant of the complex
$$
0\longrightarrow\Gamma_{\!N}\stackrel{T_N}{\longrightarrow}\dots
\longrightarrow\Gamma_{\!2}\stackrel{T_2}{\longrightarrow}\Gamma_{\!1}
\stackrel{T_1}{\longrightarrow}\Gamma{\longrightarrow}0\,.
$$
Therefore the preceding theorems describe transformation-- and
invariance--properties of determinants of complexes.

Finally we note that the partition function (\ref{2.24}) has encoded in it the
data characterising the vectorspaces associated with the resolvent
$R(S)\,$: The dimensions of $ker(S)\,,\,\Gamma\,,\,\Gamma_{\!k}\,,\,
ker(T_k)\,,\,Im(T_k)\,$ and $H^k(R(S))$ can be recovered from the
scaling behavior of this partition function under the scalings
$T_k\to{\lambda}T_k\,,\,\Phi_k\to\mu\Phi_k$ for $k=0,1,\dots,N\,$.
Note that if we had constructed the partition function by simply discarding
the divergent volume $V(ker(T))$ in (\ref{1.3}) then the resulting expression
would not encode the dimensions of $\Gamma$ and of $ker(S)=ker(T)\,$.

\vspace {1ex}

\subsection{The partition function of an arbitrary quadratic functional
scaled by a complex-valued parameter}

In the following we extend the method described
above to evaluate the partition function (\ref{1.1}) in the
general case where the functional $S$ is not required to be positive,
and where $\beta$ is an arbitrary complex-valued parameter.
Note first that if $S(\omega)=\langle\omega,
T\omega\rangle_0$ takes negative as well as positive
values then we no longer have $\ker(T)=S^{-1}(0)$
as in the preceding. However, from (\ref{2.3}) it
is clear that $\ker(T)\subseteq\Gamma$ is
independent of the choice of inner product
$\langle\cdot\,,\cdot\rangle_0$ in $\Gamma$ and by an
abuse of notation we will continue to
denote this space by $\ker(S)$. With this
notation a resolvent of an arbitrary quadratic functional
is defined as in (\ref{2.5}) above.

We decompose $ker(T)^{\perp}=\Gamma_{\!+}\oplus\Gamma_{\!-}$
such that $\widetilde{T}$ restricts to a strictly positive map
$\widetilde{T}_+\,\colon\,\Gamma_{\!+}\longrightarrow\Gamma_{\!+}$ and
a strictly negative map $\widetilde{T}_-\,\colon\,\Gamma_{\!-}
\longrightarrow\Gamma_{\!-}\,$. Then the expression (\ref{1.2}) for the
partition function
can be formally written as
\begin{equation}
Z(\beta)
=V(ker(S))\,\Bigl(\,\int_{\Gamma_{\!+}}{\cal D\/}\omega_{\!+}\,e^{-\langle
\omega_{\!+},\,\beta\widetilde{T}_{\!+}
\omega_{\!+}\rangle_{\!0}}\Bigr)\Bigl(\,\int
_{\Gamma_{\!-}}{\cal D\/}\omega_{\!-}\,
e^{-\langle\omega_{\!-},\,\beta\widetilde{T}_{\!-}
\omega_{\!-}\rangle_{\!0}}\Bigr) \label{2.41}
\end{equation}
We evaluate the integrals $\int_{\Gamma_{\!\pm}}(\cdots)$ in (\ref{2.41}) as
follows. The maps $\widetilde{T}_+$ and $-\widetilde{T}_-$
are strictly positive, so
$\beta\widetilde{T}_{\pm}=(\pm\beta)(\pm\widetilde{T}_{\pm})$ is strictly
positive for $\beta\in{\bf R_{\pm}}\,$. In this case
\begin{equation}
\int_{\Gamma_{\pm}}{\cal D\/}\omega_{\!{\pm}}\,
e^{-\langle\omega_{\!{\pm}},\,\beta
\widetilde{T}_{\!{\pm}}\omega_{\!{\pm}}
\rangle_{\!0}}=\pi^{d_{\pm}/2}\,(\pm\beta)
^{d_{\pm}/2}\,det(\pm\widetilde{T}_{\!{\pm}})^{-1/2} \label{2.42}
\end{equation}
where $d_{\pm}=dim\Gamma_{\!\pm}\,$. We extend (\ref{2.42})
to $\beta\in{\bf C}$
by analytic continuation; to do this we must fix a convention for defining
$z^a$ for $z\in{\bf C}$ and $a\in{\bf R}\,$. The natural way to do this is
to write $z=|z|e^{i\theta}$ with $\theta\in[-\pi\,,\pi]$ and set
$z^a=|z|^ae^{i{\theta}a}\,$. This is well-defined for all $a\in{\bf R}$
provided $z\not\in{\bf R_-}\,$; if $z\in{\bf R_-}\;,\;\,z{\ne}0$ there is
a phase ambiguity. Now from
(\ref{2.42}) we get the following expression for (\ref{2.41}):
\begin{equation}
Z(\beta)
=\pi^{\zeta/2}\,\beta^{d_+/2}\,(-\beta)^{d_-/2}\,det(|\widetilde{T}|)
^{-1/2}\,V(ker(S)) \label{2.43}
\end{equation}
where $|\widetilde{T}|=\sqrt{\widetilde{T}^2}$ is defined via spectral theory.
Let ${\bf C_+}$ and ${\bf C_-}$ denote the upper and lower halfplanes
of ${\bf C}$ respectively. We write $\beta=|\beta|e^{i\theta}$ with
$\theta\in[-\pi,\pi]\,$; then for $\beta\in{\bf C_{\pm}}$ we have
$-\beta=|\beta|e^{i(\theta\mp\pi)}$ with $\theta\mp\pi\in\;$ \break
$[-\pi,\pi]\,$.
Using $\zeta=d_++d_-\,$, setting $\eta=d_+-d_-$ and formally evaluating
and normalising the divergent volume $V(ker(S))$ from a resolvent $R(S)$ as
previously, we obtain from (\ref{2.43}) the following expression for the
partition function:
\begin{eqnarray}
\lefteqn{Z(\beta\,;R(S),\langle\cdot\,,\cdot\rangle_H,\langle\cdot\,,
\cdot\rangle)}\nonumber \\
&=&\pi^{\zeta/2}e^{-\frac{i\pi}{4}((\frac{2\theta}{\pi}\mp1)\zeta\;\pm\;\eta)}
\,|\beta|^{-\zeta/2}\,\widetilde{Z}(R(S),\langle\cdot\,,\cdot\rangle_H,
\langle\cdot\,,\cdot\rangle)\label{2.44}
\end{eqnarray}
where
\begin{eqnarray}
\lefteqn{\widetilde{Z}(R(S),\langle\cdot\,,\cdot\rangle_H,
\langle\cdot\,,\cdot\rangle)}\nonumber \\
&=&det(|\widetilde{T}|)^{-1/2}\prod_{k=0}^N\Bigl(det(\widetilde{T}_{k+1}^*
\widetilde{T}_{k+1})^{1/2}\,det(\Phi_k^*\Phi_k)^{-1/2}\Bigr)^{(-1)^k}
\,.\label{2.45}
\end{eqnarray}
The partition function
(\ref{2.44}) is well-defined for $\beta\in{\bf C}-{\bf R}\,$;
for $\beta\in{\bf R}$ there is a phase ambiguity. Note that
\begin{equation}
\zeta=\zeta(0\,|\,|T|)\;\;,\;\;\;\eta=\eta(0\,|\,T) \label{2.46}
\end{equation}
where $\zeta(s\,|\,|T|)$ and $\eta(s\,|\,T)$ are the zeta- and eta-functions
of $|T|$ and $T$ respectively. In the infinite-dimensional case considered in
\S3 (\ref{2.46}) will be used to regularise the quantities $\zeta$ and $\eta$
in the partition function (\ref{2.44}).
By analogy with (\ref{2.28})
we can consider the partition function
as an element $\widehat{Z}(\beta\,;R(S),
\langle\cdot\,,\cdot\rangle)$
in $\otimes_{k=0}^N\Lambda^{max}H^k(R(S))^{*^{k+1}}\,$.

There are 2 special cases of particular interest: Setting $\beta=1$ in
(\ref{2.44}) gives
\begin{equation}
Z(R(S),\langle\cdot\,,\cdot\rangle_H,\langle\cdot\,,\cdot\rangle)=
\pi^{\zeta/2}e^{\pm\frac{i\pi}{4}(\zeta-\eta)}\widetilde{Z}(R(S),
\langle\cdot\,,\cdot\rangle_H,\langle\cdot\,,\cdot\rangle)\,. \label{2.47}
\end{equation}
The case of relevance to the partition functions in the semiclassical
approximation (\ref{1.14}) is $\beta=i\lambda\;,\;\lambda\in{\bf R}\,$.
In this case we have $\theta=\pm\pi/2$ for $\lambda\in{\bf R}_{\pm}$ and the
partition function (\ref{2.44}) is
\begin{equation}
Z(i\lambda\,;R(S),\langle\cdot\,,\cdot\rangle_H,\langle\cdot\,,\cdot\rangle)=
\pi^{\zeta/2}\,e^{\mp\frac{i\pi}{4}\eta}\,\lambda^{-\zeta/2}\,
\widetilde{Z}(R(S),\langle\cdot\,,\cdot\rangle_H,\langle\cdot\,,
\cdot\rangle)\,. \label{2.48}
\end{equation}

Note that $\zeta$ and $\eta$ depend only on
the functional $S$, and not on the choice of inner product
in $\Gamma$. Therefore the transformation-- and
invariance--properties of the partition function described in the preceding
continue to hold for the general partition function (\ref{2.44}).

Finally we point out that the methods
and results of this section also apply to real-valued
quadratic functionals on complex vectorspaces.
(Then for any choice of complex
inner product $\langle\cdot\,,\cdot\rangle$
in $\Gamma$ we can write $S(\omega)=\langle
\omega,T\omega\rangle\,$, with $T$ selfadjoint, as before).
Since the integration in (\ref{1.1})
in this case is over the real vectorspace
underlying $\Gamma\,$, which has twice the
dimension of $\Gamma\,$, the expressions for
the partition functions in this case are
the squares of those above.

\section{The partition function in the infinite-dimensional case}

In this \S\ we extend the construction of the
partition function in \S2 to the following
infinite-dimensional setup and derive its variation under variation of the
inner products and maps in the resolvent by the group actions introduced
in \S2. In doing so we extend results obtained by Schwarz in \cite{S}.
It turns out that a version of theorem 2.6 from \S2 continues to hold in
the following infinite-dimensional case, determining a new invariance-property
of the partition function.
We will use some standard results
in elliptic operator theory --for the proofs see e.g. \cite{Gi}.

\subsection{Construction of the partition function}

Throughout the following $M$ denotes a compact
oriented riemannian manifold without boundary.
The functionals we consider are defined on the space
$\Gamma=\Gamma(M,\xi)$
of smooth sections in a real Hermitian
vectorbundle\footnote{Complex bundles can
also be treated, in the same way as complex vectorspaces were treated in \S2.}
$\xi$ over $M$.
Given a metric on $M$ and a Hermitian structure in $\xi$, i.e. an inner
product $\langle\cdot\,,\cdot\rangle_x$ in
each fibre $\xi_x\,$ smoothly varying with $x{\in}M$,
we obtain an inner product in $\Gamma$ :
\begin{equation}
\langle\omega,\tau\rangle_0
=\int_M\langle\omega(x),\tau(x)\rangle_x\,vol(x) \label{3.1}
\end{equation}
where $vol$ is the volume form on $M$ determined by the metric
and orientation. Our evaluation of the
partition function of a quadratic functional
$S(\omega)$ on $\Gamma$ from a resolvent $R(S)$ of
$S$ (as in (\ref{2.5})) requires the following conditions to be satisfied.

(i) The functional can be written as $S(\omega)
=\langle\omega,T\omega\rangle_0$ where $T$ is a
formally self-adjoint differential operator of order $d>0$ on $\Gamma\,$.

(ii) The resolvent $R(S)$ in
(\ref{2.5}) is of the following type : Each space $\Gamma_{\!k}$
is the space $\Gamma_{\!k}=\Gamma(M,\xi_k)$ of smooth sections in a
real Hermitian vectorbundle $\xi_k$ over $M$, and
each map $T_k\,\colon\,\Gamma_{\!k}\longrightarrow
\Gamma_{\!k-1}$ is a differential operator of the same order $d$ as
$T$ in (i).

(iii) The following chain of maps determined by the resolvent $R(S)$ is an
elliptic complex:
\begin{equation}
0\stackrel{0}{\longrightarrow}\Gamma_{\!N}
\stackrel{T_N}{\longrightarrow}\dots\longrightarrow
\Gamma_{\!1}\stackrel{T_1}{\longrightarrow}
\Gamma\stackrel{T}{\longrightarrow}\Gamma\stackrel
{T_1^*}{\longrightarrow}\Gamma_{\!1}\longrightarrow
\dots\stackrel{T_N^*}{\longrightarrow}
\Gamma_{\!N}\stackrel{0}{\longrightarrow}0\,. \label{3.2}
\end{equation}
Condition (iii) is equivalent to requiring that the Laplace-operators
\begin{equation}
\Delta_k=T_k^*T_k+T_{k+1}T_{k+1}^*\,\colon\,
\Gamma_{\!k}\longrightarrow\Gamma_{\!k} \label{3.3}
\end{equation}
are elliptic of order $2d$ for all $k=0,1,\dots,N$
(with $\Gamma_{\!0}=\Gamma\,,\,T_0=T\,,\,
T_{\!N+1}=0$).
The adjoint maps $T_k^*$ in (\ref{3.2})
are obtained by choosing a Hermitian structure in
each $\xi_k$ and a metric on $M$ and defining
an inner product $\langle\cdot\,,\cdot\rangle_k$
in each $\Gamma_k$ by analogy with (\ref{3.1}). Although the complex
(\ref{3.2}) depends on the choice of
Hermitian structures and metric, it is easy to see that
the ellipticity of (\ref{3.2}), or lack
of it, is independent of these choices.
We denote the inner products $\langle\cdot\,,\cdot
\rangle_N,\dots,\langle\cdot\,,\cdot\rangle_0$
collectively by $\langle\cdot\,,\cdot\rangle$
as previously.

When the functional $S$ satisfies (i) and the resolvent
$R(S)$ satisfies (ii) and (iii) we say that
it is an elliptic resolvent of $S$. In this case it
follows from Hodge-theory that the cohomology
spaces (\ref{1.5}) of $R(S)$ are finite-dimensional.
Given an elliptic resolvent $R(S)$ for $S$ we equip the cohomology spaces
$H^k(R(S))$ with inner products $\langle\cdot\,,\cdot\rangle_{H^k}\,$, then
with $\beta=
|\beta|e^{i\theta}\in{\bf C_{\pm}}$ the partition function of $S$
can be formally evaluated as in \S2 to
get the formal expression (\ref{2.44}) :
\begin{eqnarray}
\lefteqn{Z(\beta\,;R(S),\langle\cdot\,,\cdot\rangle_H,
\langle\cdot\,,\cdot\rangle)}\nonumber \\
&=&\pi^{\zeta/2}\,
e^{-\frac{i{\pi}}{4}((\frac{2\theta}{\pi}\mp1)\zeta\,\pm\,\eta)}\,
|\beta|^{-\zeta/2}\,
\widetilde{Z}(R(S),\langle\cdot\,,\cdot\rangle_H,
\langle\cdot\,,\cdot\rangle)\label{3.4}
\end{eqnarray}
with
\begin{equation}
\widetilde{Z}(R(S),\langle\cdot\,,\cdot\rangle_H,\langle\cdot\,,\cdot\rangle)
=\widetilde{Z}(R(S),\langle\cdot\,,\cdot\rangle){\cdot}
\Psi(R(S),\langle\cdot\,,\cdot\rangle_H,\langle\cdot\,,\cdot\rangle)
\label{3.5}
\end{equation}
where
\begin{equation}
\widetilde{Z}(R(S),\langle\cdot\,,\cdot\rangle)
=\det(\widetilde{T}^2)^{-1/4}\,\prod_{k=1}^N\det(\widetilde{T}_k^*
\widetilde{T}_k)^{\frac{1}{2}(-1)^{k-1}} \label{3.6}
\end{equation}
and
\begin{equation}
\Psi(R(S),\langle\cdot\,,\cdot\rangle_H,\langle\cdot\,,\cdot\rangle)
=\prod_{k=0}^N\det(\Phi_k^*\Phi_k)^{\frac{1}{2}(-1)^{k+1}} \label{3.7}
\end{equation}
The expressions (\ref{3.5}), (\ref{3.6}) and
(\ref{3.7}) follow from (\ref{2.45}).
The expression (\ref{3.4}) for
the partition function is not well-defined
a priori because it involves the determinants of operators
on infinite-dimensional vectorspaces, and
because the quantities $\zeta$ and $\eta$ do not have
well-defined finite values in the present infinite-dimensional case.

To obtain a well-defined expression for the partition function
(\ref{3.4}) we
use zeta-function regularisation techniques, following \cite{S}.
Note that in the present case the maps $\Phi_k$ continue to map between
finite-dimensional vectorspaces so $\Psi$ in (\ref{3.7}) is well-defined.
Given a choice of orientations for the cohomology spaces $H^k(R(S))$ then
considered as a functional of $\langle\cdot\,,\cdot\rangle_H\,$ $\Psi$ can
be interpreted as an element
\begin{equation}
\widehat{\Psi}(R(S),\langle\cdot\,,\cdot\rangle)=
\otimes_{k=0}^N\Lambda^{max}H^k(R(S))^{*^{k+1}}\label{3.7a}
\end{equation}
as previously in (\ref{2.26}).
The same is therefore true of the zeta-regularised partition function
which we will obtain from (\ref{3.4}).
Remarkably, it turns out that the zeta-regularised
expression for the partition function has
analogous symmetry properties to those established
for the finite-dimensional case
in \S2 under variation of the inner products and resolvent :
We will derive generalisations of the formulae
(\ref{2.32}) and (\ref{2.40}) for the variation of the partition
function under variations of the inner
products in the spaces $\Gamma_{\!k}$ resulting from
variations of the Hermitian structures and metric on $M$, and under
variations of the resolvent of the type considered in \S2.
It turns out that when the dimension of the manifold is odd the
modulus of the partition function is invariant
under {\it all} the mentioned variations.
This extends results of Schwarz in \cite{S} obtained in the case where
the cohomology of the resolvent vanishes.

The zeta-regularisation techniques and subsequent formulae that we derive
are based on
some standard results concerning the zeta-functions of the
Laplacians $\Delta_k$ in (\ref{3.3}) and of $T_k^*T_k\,$. For the sake of
completeness we give a brief summary of these.
Since the complex (\ref{3.2}) is elliptic the Laplace-operators
(\ref{3.3}) are positive, selfadjoint elliptic
differential operators. Let $\lbrace\mu_j^{(k)}\rbrace$
denote the non-zero eigenvalues of
$\Delta_k$, with $\mu_j^{(k)}{\le}\mu_{j+1}^{(k)}$ for all $j$, and
each $\mu_j^{(k)}$ appearing the same number of
times as its multiplicity. The zeta-function for $\Delta_k$ is
\begin{equation}
\zeta(s\,|\,\Delta_k)=\sum_j\frac{1}{(\mu_j^{(k)})^s}\,. \label{3.8}
\end{equation}
This is well-defined and analytic for $s{\in}{\bf C}$
with Re$(s)>>0$; it extends by analytic
continuation to a meremorphic
function on ${\bf C}$, regular at $s=0$.
{}From the property $T_k{\circ}T_{k+1}=0$ of the resolvent
it follows that $(T_k^*T_k){\circ}(T_{k+1}T_{k+1}^*)=
(T_{k+1}T_{k+1}^*){\circ}(T_k^*T_k)=0$,
and therefore $\lbrace\mu_j^{(k)}\rbrace$ is the
union of the non-zero eigenvalues of $T_k^*T_k$
and $T_{k+1}T_{k+1}^*$. Since $T_{k+1}T_{k+1}^*$
has the same non-zero eigenvalues as
$T_{k+1}^*T_{k+1}$ (because if $T_{k+1}^*T_{k+1}
(\omega)=\lambda\omega$
then $\,T_{k+1}T_{k+1}^*
(T_{k+1}\omega)\,$
$={\lambda}(T_{k+1}\omega)$, etc.) it follows that
\begin{equation}
\lbrace\mu_j^{(k)}\rbrace=\lbrace\lambda_l^{(k)}
\rbrace\,\cup\,\lbrace\lambda_m^{(k+1)}\rbrace
\label{3.9}
\end{equation}
where $\lbrace\lambda_l^{(k)}\rbrace$ and
$\lbrace\lambda_m^{(k+1)}\rbrace$ are the non-zero
eigenvalues of $T_k^*T_k$ and $T_{k+1}^*T_{k+1}$
respectively. We will need the following

\vspace {1ex}
\noindent {\bf Lemma 3.1.} {\it
The zeta-functions $\zeta(s\,|\,T_k^*T_k)$ (defined by analogy with
(\ref{3.8})) are well-defined and
analytic for Re$(s)>>0$ for all $k=0,1,\dots,N$.
They extend by analytic continuation to
meremorphic functions on ${\bf C}$,
which are regular at $s=0$, and satisfy the formulae}
\begin{equation}
\zeta(s\,|\,\Delta_k)=\zeta(s\,|\,T_k^*T_k)+\zeta(s\,|\,T_{k+1}^*T_{k+1})\,.
\label{3.10}
\end{equation}

\vspace {1ex}
\noindent {\it Proof.}
Since all the $\mu_j^{(k)}$ and $\lambda_l^{(k)}$
are positive it follows from (\ref{3.9}) and
the fact that (\ref{3.8}) is well-defined for
Re$(s)>>0$ that the $\zeta(s\,|\,T_k^*T_k)$ are
well-defined for Re$(s)>>0$. It now follows from (\ref{3.9})
that the formula (\ref{3.10}) holds
for Re$(s)>>0$. A simple induction argument based on
(\ref{3.10}) with Re$(s)>>0$ and starting
with $\zeta(s\,|\,\Delta_N)=\zeta(s\,|\,T_N^*T_N)$
(obtained by setting $k=N$ in (\ref{3.10})) shows
that the $\zeta(s\,|\,T_k^*T_k)$ are analytic
for Re$(s)>>0$, and that their analytic
continuations to meremorphic
functions on ${\bf C}$ are regular at $s=0$, because the
$\zeta(s\,|\,\Delta_k)$ have these properties.

The analytic continuations of the
$\zeta(s\,|\,T_k^*T_k)$ can be used to regularise
the determinants appearing in the partition function in the
present case :
Formally we have
\begin{equation}
\det(\widetilde{T}_k^*\widetilde{T}_k)=
\prod_j\lambda_j^{(k)}=e^{-\zeta'(0\,|\,T_k^*T_k)}\,.
\label{3.11}
\end{equation}
We use the analytic continuation of
$\zeta(s\,|\,T_k^*T_k)$ to give meaning to the r.h.s. of
(\ref{3.11}) and substitute this for
$\det(\widetilde{T}_k^*\widetilde{T}_k)$ in the expression
(\ref{3.6}) above.
{}From (\ref{2.46}) we see that the quantities
$\zeta$ and $\eta$ in (\ref{3.4}) are formally given by
\begin{equation}
\zeta=\zeta(0\,|\,|T|)\;\;,\;\;\,\eta=\eta(0\,|\,T) \label{3.13}
\end{equation}
where $\zeta(s\,|\,|T|)$ and $\eta(s\,|\,T)$ are the zeta- and eta-functions
of $|T|$ and $T$ respectively.
We could also obtain
(\ref{3.13}) by repeating the formal calculations in \S2.2
with the determinants replaced by their zeta-regularisations (defined by
analogy with (\ref{3.11})).
Theorems 3.2 and 3.3 below enable us to give
finite, well-defined meaning to $\zeta$ and $\eta$
in (\ref{3.13}) via the analytic
continuations of $\zeta(s\,|\,|T|)$ and $\eta(s\,|\,T)$ to $s=0$.
We substitute these values into (\ref{3.4}) and thus obtain a finite,
well-defined expression for the partition function
(\ref{3.4}). To formulate the theorems we need
the following standard result in elliptic
operator theory: There exists $\epsilon>0$ and
$\delta>0$ such that the following heat kernel
expansion holds for $0<t<\delta\,$:
\begin{equation}
\mbox{Tr}(e^{-t\Delta_k})=\sum_{0{\le}l{\le}l_0}a_l(\Delta_k)t^{-l}
+O(t^{\epsilon}) \label{3.14}
\end{equation}

\vspace {1ex}
\noindent {\bf Theorem 3.2.}
{\it
The zeta-function $\zeta(s\,|\,|T|)$ is well-defined
and analytic for Re$(s)>>0$. Its analytic
continuation to a meremorphic function
on ${\bf C}$ is regular at $s=0$ and satisfies the formula
\begin{equation}
\zeta=\zeta(0\,|\,|T|)=\sum_{k=0}^N(-1)^k(a_0(\Delta_k)-dimH^k(R(S)))
\label{3.15}
\end{equation}
where $a_0(\Delta_k)$ is the coefficient corresponding to $l=0$ in
(\ref{3.14}).}

\vspace {1ex}
\noindent {\it Proof.}
Let $\lbrace\lambda_j\rbrace$ denote the non-zero eigenvalues of $T\,$;
from the definition of the zeta-function we have
$$
\zeta(2s\,|\,|T|)=\sum_j\frac{1}{|\lambda_j|^{2s}}=
\sum_j\frac{1}{(\lambda_j^2)^s}=\zeta(s\,|\,T^2)
$$
Since $T^2=T_0^*T_0$ it follows from lemma 3.1 that $\zeta(s\,|\,|T|)
=\zeta(\frac{s}{2}\,|\,T_0^*T_0)$ is
well-defined and analytic for Re$(s)>>0$,
with analytic continuation regular at $s=0$. A simple
induction argument based on
(\ref{3.10}) and starting with $\zeta(s\,|\,T_0^*T_0)=\zeta(s\,|\,\Delta_0)
-\zeta(s\,|\,T_1^*T_1)$ gives
$$
\zeta(s\,|\,T_0^*T_0)=\sum_{k=0}^N(-1)^k\zeta(s\,|\,\Delta_k)\,.
$$
To show (\ref{3.15}) it now suffices
to show $\,\zeta(0\,|\,\Delta_k)=a_0(\Delta_k)-dim
H^k(R(S))\,$. We show this starting with the formula \cite[p.78]{Gi}
$$
\zeta(s\,|\,\Delta_k)=\frac{1}{\Gamma(s)}\int_0^{\infty}t^{s-1}\biggl(
\,\sum_{\mu_j^{(k)}>0}
e^{-t\mu_j^{(k)}}\biggr)\,dt
$$
where $\Gamma(s\!)$ is the gamma-function.
Using $\Gamma(s\!)\!=\!\frac{1}{s}\Gamma(s+1)$ and
$\sum_{\!\mu_j^{(k)}>0}e^{-t\mu_j^{(k)}}\!=\!$ \break
$\mbox{Tr}(e^{-t\Delta_k})-dim(ker(\Delta_k))$ we get
$$
\zeta(s\,|\,\Delta_k)=\frac{s}{\Gamma(s+1)}
\int_0^{\infty}t^{s-1}\Bigl(\mbox{Tr}(e^{-t\Delta_k})
-dim(ker(\Delta_k))\Bigr)\,dt\,.
$$
It is well-known that $\int_{\delta}^{\infty}t^{s-1}
(\mbox{Tr}(e^{-t\Delta_k})-dim(ker(\Delta_k)))\,dt$
is an entire function of $s{\in}{\bf C}$
for all $\delta>0$ (this follows for example
from the results of \cite[\S1.6]{Gi}).
Decomposing the integral $\int_0^{\infty}$
above into $\int_0^{\delta}+\int_{\delta}^{\infty}$
and substituting the expansion
(\ref{3.14}) for $\mbox{Tr}(e^{-t\Delta_k})$ in the $\int_0^{\delta}$ integral
leads to
$$
\zeta(0\,|\,\Delta_k)=a_0(\Delta_k)-dim(ker(\Delta_k))
=a_0(\Delta_k)-dimH^k(R(S))\,.
$$
This completes the proof.

\vspace {1ex}
\noindent {\bf Theorem 3.3.} {\it
The eta-function $\eta(s\,|\,T)$ is well-defined and analytic for \break
Re$(s)>>0$. Its analytic
continuation to a meremorphic function
on ${\bf C}$ is regular at $s=0$. It satisfies $\eta(s\,|\,T)=
\eta(s\,|\,D)$, where $D=T+\sum_{k=1}^N(T_k+T_k^*)$
is a selfadjoint elliptic operator on
$\oplus_{k=0}^N\Gamma_{\!k}$. (Here we define $T_k=0$ on $\Gamma_{\!j}$ for
$j{\ne}k\,$.) }

\vspace {1ex}
\noindent {\it Proof.}
Since $T_k{\circ}T_{k+1}=0$ we have
\begin{equation}
D^2=\oplus_{k=0}^N\Delta_k\,,
\label{3.16}
\end{equation}
which is elliptic, hence $D$ is elliptic.
A result due to Atiyah, Patodi and Singer \cite{APS} in the case when $dimM$
is odd, and Gilkey \cite[\S4.3]{Gi} in the case when $dimM$ is even
now states that
$\eta(s\,|\,D)$ is well-defined and analytic for Re$(s)>>0$, with analytic
continuation regular at $s=0\,$. We will show that
$\eta(s\,|\,T)=\eta(s\,|\,D)$, which
proves the theorem.

We can decompose $D=\widetilde{D}\oplus\widetilde{T}$ with
$\widetilde{D}=\sum_{k=1}^N(T_k+T_k^*)$ acting on
$(\oplus_{k=1}^N\Gamma_{\!k})\oplus\ker(T)$ and $\widetilde{T}$ the
restriction of $T$ to $\ker(T)^{\perp}\,$, as previously.
{}From (\ref{3.16}) it follows that the
zeta-function of $D^2=
\widetilde{D}^2\oplus\widetilde{T}^2$ is well-defined for Re$(s)>>0\,$
, which implies that the eta-functions of $D\,,\,\,\widetilde{D}\,,$
and $\widetilde{T}$
are well-defined for Re$(s)>>0\,$. We will show
that $\eta(s\,|\,\widetilde{D})=0$ for
Re$(s)>>0\,$, from which it follows that $\eta(s\,|\,D)=\eta(s\,|\,T)\,$.
To show this it suffices to construct
an isomorphism $\omega\mapsto\omega'$
which maps each
eigenvector $\omega$ of $\widetilde{D}$ with
eigenvalue $\lambda$ to an eigenvector with eigenvalue $-\lambda\,$.
We decompose $\omega\in\oplus_{k=1}^N\Gamma_{\!k}
\oplus\ker(T)$ as $\omega=\oplus_{k=0}^N\omega_k
\,\,,\,\,\omega_k\in\Gamma_{\!k}\,$, and define the
isomorphism $\omega\mapsto\omega'$
by $\omega'=\oplus_{k=0}^N(-1)^k\omega_k\,$.
To show that this isomorphism has the
required property we consider the eigenvalue equation
$\widetilde{D}\omega=\lambda\omega\,$: It is equivalent to the collection
of equations
\begin{equation}
\lbrace\,T_k(\omega_k)+T_{k-1}^*(\omega_{k-2})=
\lambda\omega_{k-1}\,\rbrace\;\;\;\;\;\;\;\;\;
k=1,\dots,N+1 \label{3.17}
\end{equation}
(with $\omega_{-1}=0\,$). From this we calculate
\begin{eqnarray*}
T_k{\omega'}_k+T_{k-1}^*{\omega'}_{k-2}
&=&(-1)^k\,(\,T_k\omega_k+T_{k-1}^*\omega_{k-2}\,) \\
&=&-(-1)^{k-1}\lambda\omega_{k-1} \\
&=&(-\lambda){\omega'}_{k-1}\,.
\end{eqnarray*}
It follows from
(\ref{3.17}) that $\widetilde{D}\omega'=-\lambda\omega'$ as required.
This completes the proof. (The statement $\eta(0\,|\,T)=\eta(0\,|\,D)$ is
similar to \cite[(I), proposition(4$\cdot$20)]{APS}
and can also be shown by
a proof similar to the one given there).

\subsection{Variation of the partition function under
variation of inner products and resolvent}

In the following the group actions on the partition function considered in \S2
are generalised to the present case. Formulae
for the variation of the partition function
under the infinitessimal versions of these actions are derived
and resulting invariance properties of the partition function are pointed out.

In the following we assume given a metric on $M$
and Hermitian structures in the bundles
$\xi_k$ and use them to construct inner products
$\langle\cdot\,,\cdot\rangle_k$ in the spaces
$\Gamma_{\!k}$ by analogy with
(\ref{3.1}), denoted collectively by $\langle\cdot\,,\cdot\rangle\,$.
Let $GL(\xi)$ denote the group of isomorphisms of
the vectorbundle $\xi$ , i.e. $A{\in}GL(\xi)$
is a collection of isomorphisms $\lbrace\,A_x{\in}GL(\xi_x)\,
\rbrace_{x{\in}M}$ varying
smoothly with $x$. Then $A{\in}GL(\xi)$ is an
isomorphism of $\Gamma=\Gamma(M,\xi)$ defined by
$A(\omega)(x)=A_x(\omega(x))\,$.
The group $GL(\xi_N){\times}{\cdots}{\times}GL(\xi_1){\times}
GL(\xi)$ acts on the inner products $\langle\cdot\,,
\cdot\rangle$ in the spaces $\Gamma_{\!k}=
\Gamma(M,\xi_k)$ by analogy with the action described in \S2
(see (\ref{2.2})).

\vspace {1ex}
\noindent {\bf Lemma 3.4.} {\it
All inner products in the spaces $\Gamma_{\!k}$ constructed as in (\ref{3.1})
from a metric on $M$ and Hermitian structures
in the bundles $\xi_k$ can be obtained
from the group action above on the inner products constructed from the given
metric and Hermitian structures.}

\vspace {1ex}
\noindent {\it Proof.}
Let $vol$ and $vol'$ be the volume forms on
$M$ constructed from 2 different metrics, and
let $\lbrace\langle\cdot,\cdot\rangle_x
\rbrace_{x{\in}M}$ and $\lbrace\langle\cdot,\cdot
{\rangle_x}'\rbrace_{x{\in}M}$ be 2 different
Hermitian structures in $\xi_k\,$. Then
$vol'(x)=f(x)vol(x)$ and ${\langle}u,v{\rangle_x}'=
{\langle}u,Q_xv\rangle_x$ where $f$
is a smooth strictly positive function on
$M$ and $Q_x$ is a map on $(\xi_k)_x$ strictly
positive and Hermitian w.r.t. $\langle\cdot\,,
\cdot\rangle_x\,$, varying smoothly with
$x{\in}M$. Then the inner products $\langle\cdot\,,
\cdot\rangle$ and $\langle\cdot\,,\cdot
\rangle'$ in $\Gamma_{\!k}$ constructed by analogy with
(\ref{3.1}) are related by $\langle\omega,
\tau\rangle'=\langle\omega,(fQ)\tau\rangle$ where
$(Q\tau)(x)=Q_x(\tau(x))\,$. It follows that
$\langle\cdot\,,\cdot\rangle'$ is obtained from the
action of $A=(fQ)^{1/2}$ on $\langle\cdot\,,
\cdot\rangle$ (with $Q_x^{1/2}$ defined via spectral theory).

\vspace {1ex}

We will derive the variation of the modulus of the partition function
$Z(\beta\,;R(S),\langle\cdot\,,\cdot\rangle_H,\langle\cdot\,,\cdot\rangle)\,$,
given by (\ref{3.4}) under the infinitessimal version of
the above group action on the inner products $\langle\cdot\,,\cdot\rangle\,$.
We begin by deriving the variation of
$\widetilde{Z}(R(S),\langle\cdot\,,\cdot\rangle_H,
\langle\cdot\,,\cdot\rangle)\,$,
given by (\ref{3.5}), under the infinitessimal version of the group action.
To do this we need the variation of
$\widetilde{Z}(R(S),\langle\cdot\,,\cdot\rangle)\,$,
given by (\ref{3.6}); this is readily obtained from a theorem in \cite{S}.
Let $A(u)=(A_N(u),\dots,A_1(u),A_0(u))$ be a smooth
curve in $GL(\xi_N)\times\cdots{\times}
GL(\xi_1){\times}GL(\xi)$ with $A(0)=I$ (the identity).
We define the endomorphisms
$B_k{\in}End(\xi_k)$ for $k=0,1,
\dots,N$ by $B_k=\frac{d}{du}\Bigl|_{u=0}(A_k^*(u)A_k(u))\,$.
These induce linear maps
$B_k\,\colon\,\Gamma_{\!k}\longrightarrow\Gamma_{\!k}$ with the property
$$
\frac{d}{du}\biggl|_{u=0}\langle\omega,\tau\rangle_k^{(A(u))}=
\langle\omega,B_k\tau\rangle_k
$$
Let $B\in\,End(\xi_k)$ be arbitrary.
{}From elliptic operator theory we have the following
asymptotic expansion for $t{\rightarrow}0_+$ :
\begin{equation}
\mbox{Tr}(B\,e^{-t\Delta_k})
=\sum_{0{\leq}l{\leq}l_0}a_l(B\,|\,\Delta_k)\,t^{-l}+O(t^{\epsilon})
\label{3.18}
\end{equation}
for some $\epsilon>0\,$. By Hodge-theory the spaces
$$
{\cal H\/}_k=ker(T_k)\,\cap\,Im(T_{k+1})^{\perp}=ker(\Delta_k)
$$
are finite-dimensional. Let $P_{[{\cal H\/}_k]}$
and $P_{[{\cal H\/}_k^{\perp}]}$
denote the orthogonal projections
of the completion of
$\Gamma_{\!k}$ w.r.t. $\langle\cdot\,,\cdot\rangle_k$
onto ${\cal H\/}_k$ and ${\cal H\/}_k^{\perp}$ respectively.
Then we have
\begin{eqnarray*}
\mbox{Tr}(B\,e^{-t\Delta_k})
&=&\mbox{Tr}(B\,P_{[{\cal H\/}_k^{\perp}]}\,e^{-t\Delta_k})+
   \mbox{Tr}(B\,P_{[{\cal H\/}_k]}\,e^{-t\Delta_k}) \\
&=&\mbox{Tr}(B\,P_{[{\cal H\/}_k^{\perp}]}\,e^{-t\Delta_k})+
   \mbox{Tr}(B\,P_{[{\cal H\/}_k]})\,.
\end{eqnarray*}
Combining this with (\ref{3.18}) we get
the following asymptotic expansion for $t{\rightarrow}0_+$ :
\begin{equation}
\mbox{Tr}(B\,P_{[{\cal H\/}_k^{\perp}]}\,e^{-t\Delta_k})
=\sum_{0{\leq}l{\leq}l_0}a_l(B\,P_{[{\cal H\/}_k^
{\perp}]}\,|\,\Delta_k)\,t^{-l}
+O(t^{\epsilon}) \label{3.19}
\end{equation}
where
\begin{eqnarray}
a_0(B\,P_{[{\cal H\/}_k^{\perp}]}\,|\,\Delta_k)
&=&a_0(B\,|\,\Delta_k)-\mbox{Tr}(B\,P_{[{\cal H\/}_k]})
\label{3.19a} \\
a_l(B\,P_{[{\cal H\/}_k^{\perp}]}\,|\,\Delta_k)
&=&a_l(B\,|\,\Delta_k)\;\;\;\;\;\;\;\;\;\;\mbox{for}\;\;l>0\,.\nonumber
\end{eqnarray}
The following theorem is a simple reformulation of
theorem 1$'$ in \cite{S}. It expresses the variation of
$\widetilde{Z}(R(S),\langle\cdot\,,\cdot\rangle)$
under the infinitessimal action of $GL(\xi_N)\times\cdots
{\times}GL(\xi_1){\times}GL(\xi)$ on the
inner products in the spaces $\Gamma_{\!k}\,$.
($A(u)$ and $B_k$ are as above).

\vspace {1ex}
\noindent {\bf Theorem 3.5.} (Due to A. Schwartz \cite{S}).{\it
The expression $\widetilde{Z}(R(S),\langle\cdot\,,\cdot\rangle)$ given by
the zeta-regularisation of (\ref{3.6}) satisfies}
\begin{equation}
\frac{d}{du}\biggl|_{u=0}\widetilde{Z}(R(S),\langle\cdot\,,\cdot\rangle^{A(u)})
=\biggl(\,\sum_{k=0}^N\frac{1}{2}(-1)^ka_0(B_k\,
P_{[{\cal H\/}_k^{\perp}]}\,|\,\Delta_k)\biggr)\,
\widetilde{Z}(R(S),\langle\cdot\,,\cdot\rangle) \label{3.20}
\end{equation}

\vspace {1ex}
If the cohomology of the resolvent vanishes then from
(\ref{3.4}) with $\beta=1$
we see that
$\widetilde{Z}(R(S),\langle\cdot\,,\cdot\rangle)=
|Z(R(S),\langle\cdot\,,\cdot\rangle)|$ is
the modulus of the partition function of $S\,$.
In this case we obtain from theorem 3.5 the following expression
for the variation of the modulus of the partition function of $S$
under the infinitessimal group action on the inner products.

\vspace {1ex}
\noindent {\bf Corollary 3.6.} {\it
Assume the cohomology of the resolvent vanishes. Then the modulus
$|Z(R(S),\langle\cdot\,,\cdot\rangle)|$
of the partition function of $S$ with $\beta=1$ satisfies}
\begin{equation}
\frac{d}{du}\biggl|_{u=0}|Z(R(S),\langle\cdot\,,\cdot\rangle^{A(u)})|
=\biggl(\,\sum_{k=0}^N\frac{1}{2}(-1)^ka_0(B_k\,|\,
\Delta_k)\biggr)\,|Z(R(S),\langle\cdot\,,\cdot\rangle)| \label{3.21}
\end{equation}

\vspace {1ex}
The formula (\ref{3.21}) is a generalisation of the formula
(\ref{2.18}) for the finite-dimensional case.
Indeed, if the operators in the present
case were acting on finite-dimensional vectorspaces we
would have
\begin{equation}
``\;a_0(B_k\,|\,\Delta_k)=\lim_{t{\to}0}\mbox{Tr}(B_ke^{-t\Delta_k})
=\mbox{Tr}(B_k)\,.\;\mbox{''} \label{3.22}
\end{equation}
Substituting this into (\ref{3.21}) would give (\ref{2.18}).

When the dimension of $M$ is odd the modulus of the
partition function has invariance
properties which are considerably more general that in the
finite-dimensional case. This is due to the following
remarkable (although well-known)
result.

\vspace {1ex}
\noindent {\bf Theorem 3.7.} {\it
If the dimension of $M$ is odd then
$a_0(B\,|\,\Delta_k)=0$ in the asymptotic expansion
(\ref{3.18}) for all $B{\in}End(\xi_k)$ for all $k=0,1,\dots,N\,$.}

\vspace {1ex}
\noindent This is a standard result in elliptic operator theory
(see e.g. \cite[\S1.7 lemma 1.7.4(d)]{Gi}).
Combining this theorem with corollary 3.6 and lemma 3.4 leads to
the following theorem. (The statements in the theorem concerning independence
of metric and Hermitian structure were shown by Schwarz in \cite{S}).
We say that $S$ and $R(S)$ are {\it topological} if their definitions do not
depend on choices of Hermitian structures in the bundles $\xi_k$ and metric
on $M\,$.

\vspace {1ex}

\noindent {\bf Theorem 3.8.} {\it
Assume the cohomology of the resolvent vanishes.
If $dimM$ is odd then the modulus
$|Z(R(S),\langle\cdot\,,\cdot\rangle)|$ of the partition function of $S$
with $\beta=1$
is invariant under the action of $GL(\xi_N)\times\cdots{\times}
GL(\xi_1){\times}GL(\xi)$ on the inner products in the spaces $\Gamma_{\!k}$.
If $S$ and $R(S)$ are topological then $|Z(R(S),\langle\cdot\,,
\cdot\rangle)|$
is independent of the choice of the metric on $M$ and Hermitian structures
in the bundles $\xi_k\,$.}

\vspace {1ex}

\noindent {\it Proof.}
We must show that $|Z(R(S),\langle\cdot\,,\cdot\rangle^A)|=
|Z(R(S),\langle\cdot\,,\cdot\rangle)|$ for
arbitrary $A=(A_N,\dots,A_1,A_0){\in}
GL(\xi_N){\times}\cdots{\times}GL(\xi_1)
{\times}GL(\xi)\,$. For $s\in[0,1]$ set
$Q_k(s)=((1-s){\cdot}I_k+s{\cdot}A_k^*A_k)^{1/2}\,
\in\,GL(\xi_k)\,$. ( $(Q_k(s))_x$ is the
square root of a map on $(\xi_k)_x$ which is
strictly positive and selfadjoint w.r.t.
$(\langle\cdot\,,\cdot\rangle_k)_x\,$,
and is therefore well-defined via spectral theory).
Then $Q(s)=(Q_N(s),\dots,Q_1(s),$ \break
$Q_0(s))$ is a smooth curve in $GL(\xi_N)\times\cdots
{\times}GL(\xi)$ with $\langle\cdot\,,
\cdot\rangle^{Q(0)}=\langle\cdot\,,\cdot\rangle$
and $\langle\cdot\,,\cdot\rangle^{Q(1)}=\langle\cdot\,,\cdot\rangle^A\,$.
It now suffices to show that $\frac{d}{ds}
|Z(R(S),\langle\cdot\,,
\cdot\rangle^{Q(s)})|=0\,$. Set $Q(s,u)=Q(s)^{-1}Q(s+u)\,$, then
$\langle\cdot\,,\cdot\rangle^{Q(s+u)}=\langle\cdot\,,\cdot\rangle^{Q(s)Q(s,u)}=
(\langle\cdot\,,\cdot\rangle^{Q(s)})^{Q(s,u)}\,$. It follows from
(\ref{3.21}) and theorem 3.7
that
$$
\frac{d}{ds}|Z(R(S),\langle\cdot\,,\cdot\rangle^{Q(s)})|
=\frac{d}{du}\biggl|_{u=0}
|Z(R(S),(\langle\cdot\,,\cdot\rangle^{Q(s)})^{Q(s,u)})|=0\,.
$$
This completes the proof.

\vspace {1ex}
We will now show that the above symmetry properties of the modulus of the
partition function continue to hold in the general case where the
cohomology of the resolvent is not required to vanish.

\vspace {1ex}
\noindent {\bf Theorem 3.9}\hfil\break{\it
(i) The modulus of the partition function (\ref{3.4}) with $\beta=1$ satisfies
\begin{eqnarray}
\lefteqn{
\frac{d}{du}\biggl|_{u=0}|Z(R(S),\langle\cdot\,,\cdot\rangle_H,
\langle\cdot\,,\cdot\rangle^{A(u)})|}\nonumber \\
&=&\biggl(\,\sum_{k=0}^N\frac{1}{2}(-1)^ka_0(B_k\,|\,\Delta_k)\biggr)\,
|Z(R(S),\langle\cdot\,,\cdot\rangle_H,
\langle\cdot\,,\cdot\rangle)|\,. \label{3.23}
\end{eqnarray}

\noindent (ii) If $dimM$ is odd the modulus of the general partition function
(\ref{3.4}) satisfies
\begin{equation}
|Z(\beta\,;R(S),\langle\cdot\,,\cdot\rangle_H,\langle\cdot\,,\cdot\rangle)|
=|\beta|^{-\zeta/2}|Z(R(S),\langle\cdot\,,\cdot\rangle_H,
\langle\cdot\,,\cdot\rangle)| \label{3.24}
\end{equation}
with
\begin{equation}
\zeta=-\sum_{k=0}^N(-1)^kdimH^k(R(S)) \label{3.25}
\end{equation}
and is invariant under the action of $GL(\xi_N)\times\cdots{\times}GL(\xi_1)
{\times}GL(\xi)$ on $\langle\cdot\,,\cdot\rangle\,$.
It is independent of the
choice of metric on $M$ and Hermitian structures in the bundles $\xi_k\,$
when $S$ and $R(S)$ are topological.}

\vspace{1ex}
\noindent {\it Proof.} The formulae
(\ref{3.24}) and (\ref{3.25}) in (ii) follow from
the expression
(\ref{3.4}) and theorems 3.2 and 3.7. The remainder of (ii) follows
from (i) in the same way that theorem 3.8 was obtained from corollary 3.6.
The proof of (i) is based on the following formula, where
$\Psi$ is given by (\ref{3.7}):
\begin{eqnarray}
\lefteqn{\frac{d}{du}\biggl|_{u=0}\Psi(R(S),\langle\cdot\,,\cdot\rangle_H,
\langle\cdot\,,\cdot\rangle^{A(u)})}\nonumber \\
&=&\biggl(\,\sum_{k=0}^N\frac{1}{2}(-1)^k\mbox{Tr}
(P_{[{\cal H\/}_k]}B_k)\biggr)\,\Psi(R(S),\langle\cdot\,,\cdot\rangle_H,
\langle\cdot\,,\cdot\rangle)\,.\label{3.26}
\end{eqnarray}
Combining this with theorem 3.5 and
(\ref{3.19a}) gives (\ref{3.23}), proving the theorem.
The following derivation of
(\ref{3.26}) is inspired by, and modeled on
the derivation of \cite[formula (3.4)]{RS2}. We set
$$
X(u)=(\Phi_{k(u)}^{-1})^{*(u)}\Phi_{k(u)}^{-1}\,\colon\,
H^k(R(S))\stackrel{\simeq}{\longrightarrow}H^k(R(S))\,,
$$
where $\Phi_{k(u)}\,\colon\,{\cal H\/}_{k(u)}\stackrel{\simeq}
{\longrightarrow}H^k(R(S))$ is defined as in
(\ref{2.23}) and (\ref{2.24}) with the
inner product $\langle\cdot\,,\cdot\rangle_k$ in $\Gamma_{\!k}$ replaced
by $\langle\cdot\,,\cdot\rangle_k^{A_k(u)}\,$, and $(\cdot)^{*(u)}$
denotes the adjoint map
determined by $\langle\cdot\,,\cdot\rangle_k^{A_k(u)}$
and $\langle\cdot\,,\cdot\rangle_{H^k}\,$. To show (\ref{3.26}) it suffices to
show that
\begin{equation}
\frac{d}{du}\biggl|_{u=0}detX(u)
=\mbox{Tr}(P_{[{\cal H\/}_k]}B_k)\,det(X(0))\,.
\label{3.27}
\end{equation}
We see by differentiating the formula $log\,det(X)=\mbox{Tr}(log(X))$ at
$u=0$ that (\ref{3.27}) is equivalent to
\begin{equation}
\mbox{Tr}\Bigl(X(0)^{-1}X'(0)\bigr)=
\mbox{Tr}\Bigl(P_{[{\cal H\/}_k]}B_k\Bigr)\,. \label{3.28}
\end{equation}
Pick a basis $(h_j)$ for $H^k(R(S))\,$, orthonormal w.r.t. $\langle\cdot\,,
\cdot\rangle_{H^k}\,$, and set $f_j(u)=\Phi_{k(u)}^{-1}(h_j)\,$.
Let $(\eta_j)$ be a basis for ${\cal H\/}_k\,$, orthonormal w.r.t.
$\langle\cdot\,,\cdot\rangle_k\,$, and define the matrix $Y$ by
$f_j(0)=\sum_rY_{jr}\eta_r\,$. Then
\begin{eqnarray}
X_{jl}&\equiv&{\langle}h_j\,,X(0)h_l\rangle_{H^k}={\langle}f_j(0)\,,f_l(0)
\rangle_k\nonumber \\
&=&\sum_{r,s}{\langle}Y_{jr}\eta_r\,,Y_{ls}\eta_s\rangle_k
=\sum_rY_{jr}Y_{lr}=(YY^*)_{jl}\,.\label{3.29}
\end{eqnarray}
Defining the matrices $\,\;X_{jl}^{-1}\;\,$ and $\,\;X'(0)_{jl}\;\,$ by
${\langle}\;h_j\,\;,\;X(0)^{-1}h_l\;\rangle_{H^k}$ and \break
${\langle}h_j\,,X'(0)h_l
\rangle_{H^k}$ respectively it follows from (\ref{3.29}) that
\begin{eqnarray*}
\mbox{Tr}\Bigl(X(0)^{-1}X'(0)\Bigr)
&=&\sum_{j,l}X_{jl}^{-1}X'(0)_{jl}
=\sum_{j,l}(YY^*)_{jl}^{-1}X'(0)_{jl} \\
&=&\sum_{j,l,r}Y_{rj}^{-1}Y_{rl}^{-1}X'(0)_{jl}\,.
\end{eqnarray*}
Also, we have
$$
\mbox{Tr}\Bigl(P_{[{\cal H\/}_k]}B_k\Bigr)=\sum_r\langle\eta_r\,,
B_k\eta_r\rangle_k=\sum_{r,j,k}Y_{rj}^{-1}Y_{rl}^{-1}{\langle}f_j(0)\,,
B_kf_l(0)\rangle_k\,.
$$
Therefore, to show (\ref{3.28}) it suffices to show that
\begin{equation}
X'(0)_{jl}={\langle}f_j(0)\,,B_kf_l(0)\rangle_k \label{3.30}
\end{equation}
We calculate
\begin{eqnarray}
X'(0)_{jl}&=&\frac{d}{du}\biggl|_{u=0}{\langle}h_j\,,X(u)h_l\rangle_{H^k}
=\frac{d}{du}\biggl|_{u=0}{\langle}f_j(u)\,,f_l(u)\rangle_k^{A_k(u)}
\nonumber \\
&=&\lim_{u{\to}0}\,\frac{1}{u}\Bigl({\langle}f_j(u)\,,f_l(u)\rangle_k^{A_k(u)}
-{\langle}f_j(0)\,,f_l(0)\rangle_k\Bigr)\,.\label{3.31}
\end{eqnarray}
Now, since
$$
ker(T_k)=Im(T_{k+1})\oplus{\cal H\/}_k=Im(T_{k+1})\oplus_{A_k(u)}{\cal H\/}
_{k(u)}\,,
$$
where $\oplus_{A_k(u)}$ denotes direct sum orthogonal w.r.t. $\langle\cdot\,,
\cdot\rangle_k^{A_k(u)}\,$, we can uniquely write
$$
f_j(u)=f_j(0)+w_j(u)\;\,,\,\;w_j(u){\in}Im(T_{k+1})
$$
and get
$$
{\langle}f_j(u)\,,f_l(u)\rangle_k^{A_k(u)}={\langle}f_j(0)\,,f_l(u)\rangle
_k^{A_k(u)}={\langle}f_j(0)\,,A_k^*(u)A_k(u)f_l(u)\rangle_k
$$
and
$$
{\langle}f_j(0)\,,f_l(0)\rangle_k={\langle}f_j(0)\,,f_l(u)\rangle_k\,.
$$
Substituting this into (\ref{3.31}) gives
$$
X'(0)_{jl}=\lim_{u{\to}0}\,\frac{1}{u}\Bigl({\langle}f_j(0)\,,
(A_k^*(u)A_k(u)-I)f_l(u)\rangle_k\Bigr)
={\langle}f_j(0)\,,B_kf_l(0)\rangle
$$
proving (\ref{3.30}) and thereby (\ref{3.26}).
This completes the proof.

\vspace{1ex}

Note from (\ref{3.25}) that when $dimM$ is odd
the quantity $\zeta=\zeta(0\,|\,T)$
in the expression
(\ref{3.4}) for the partition function is independent of metric
on $M$ and Hermitian structure in $\xi\,$. However, the quantity $\eta=
\eta(0\,|\,T)$ in
(\ref{3.4}), and therefore the phase of the partition function,
does depend on the metric and Hermitian structure in general.
(This was shown explicitly by Witten in \cite[\S2]{W} for partition functions
in the semiclassical approximation for the Witten-invariant).
For the class of quadratic functionals considered in \S4 we will apply
the Atiyah-Patodi-Singer index theorem to derive a formula involving
$\eta\,$.

{}From (\ref{3.22}) we see that (\ref{3.23}) is a generalisation to the present
infinite-dimensional case
of the formula (\ref{2.32}) for
the finite-dimensional case.

We now extend the group action on the pair
$(R(S),\langle\cdot\,,\cdot\rangle_{H(R(S))})$
considered in \S2 to the present case and
derive the variation of the partition function under the infinitessimal
version of this action. In the present case we take the group to be
$GL(\xi_N)\times\cdots{\times}
GL(\xi_1)\,$. Each $C_k{\in}GL(\xi_k)$
induces a linear map on $\Gamma_{\!k}$ which we also
denote by $C_k\,$.
The right group action of $GL(\xi_N)\times\cdots{\times}GL(\xi_1)$
on the set of resolvents $R(S)$ and collection $\langle\cdot\,,\cdot\rangle_H$
of inner products in the cohomology spaces $H(R(S))$ is defined in the same
way as in \S2. It is straightforward to see that the subset consisting of
elliptic resolvents is invariant under this action. (This follows from
lemma 10 in \cite{S}).
In the following we restrict our considerations to this subset.
The group $GL(\xi_N)\times\cdots{\times}GL(\xi_1)$ also
acts on the inner products
$\langle\cdot\,,\cdot\rangle$ in the spaces $\Gamma_{\!k}$ as before.
(The action on $\langle\cdot\,,\cdot\rangle_0$ is trivial).

A version of
theorem 2.6 continues to hold in the present infinite- \break
dimensional case.
Indeed, the calculations leading to
(\ref{2.35}) and (\ref{2.37}) continue to hold
and it follows that $(T_k^C)^{*(C)}T_k^C=C_k^{-1}T_k^*T_kC_k$ and
$(\Phi_k^C)^{*(C)}\Phi_k^C=C_k^{-1}\Phi_k^*\Phi_kC_k$ have the same
eigenvalues (with same multiplicities) as $T_k^*T_k$ and
$\Phi_k^*\Phi_k$ respectively. Therefore the zeta-functions of
$(T_k^C)^{*(C)}T_k^C$ and \break
$(\Phi_k^C)^{*(C)}\Phi_k^C$ are equal to those
of $T_k^*T_k$ and $\Phi_k^*\Phi_k$ respectively, so their
zeta-regularised determinants are the same.
Since $\langle\cdot\,,\cdot\rangle_0$ is unchanged by the action of
$GL(\xi_N)\times\cdots{\times}GL(\xi_1)$ on $\langle\cdot\,,\cdot\rangle$
the map $\widetilde{T}$ is unchanged,
so from (\ref{3.4})--(\ref{3.7}) we get

\vspace{1ex}

\noindent {\bf Theorem 3.10.} {\it
For all $C=(C_N,\dots,C_1){\in}GL(\xi_N)\times\cdots{\times}
GL(\xi_1)$ the partition function (\ref{3.4}) satisfies}
\begin{equation}
Z(\beta\,;R(S)^C,\langle\cdot\,,\cdot\rangle_H^C,
\langle\cdot\,,\cdot\rangle^C)=
Z(\beta\,;R(S),\langle\cdot\,,\cdot\rangle_H,\langle\cdot\,,\cdot\rangle)
\label{3.32}
\end{equation}

\vspace{1ex}

The variation of $Z(\beta\,;R(S),\langle\cdot\,,\cdot\rangle_H,
\langle\cdot\,,\cdot\rangle)$ under the variation of the pair
$(R(S),\langle\cdot\,,\cdot\rangle_H)$ by the above group action can
now be obtained by combining theorem 3.10 with theorem 3.9 and noting
that $\zeta$ and $\eta$ in (\ref{3.4}) are independent of the choice of
resolvent $R(S)\,$. Let $C(u)$ be a smooth curve in $GL(\xi_N)\times
\cdots{\times}GL(\xi_1)$ with $C(0)=I$ and set $D_k=\frac{d}{du}\Bigl|_
{u=0}(C_k^*(u)C_k(u))\,$. The induced linear maps on $\Gamma_{\!k}$ are
also denoted by $D_k\,$.

\vspace{1ex}

\noindent {\bf Corollary 3.11.} {\it
The partition function (\ref{3.4}) satisfies}
\begin{eqnarray}
\lefteqn{\frac{d}{du}\biggl|_{u=0}Z(\beta\,;R(S)^{C(u)},\langle\cdot\,,
\cdot\rangle_H^{C(u)},\langle\cdot\,,\cdot\rangle)}\nonumber \\
&=&\biggl(\sum_{k=1}^N\frac{1}{2}(-1)^{k+1}a_0(D_k\,|\,\Delta_k)\biggr)\,
Z(\beta\,;R(S),\langle\cdot\,,\cdot\rangle_H,\langle\cdot\,,\cdot\rangle)\,.
\label{3.33}
\end{eqnarray}
{\it
If $dimM$ is odd the partition function is invariant under the group action
of $GL(\xi_N)\times\cdots{\times}GL(\xi_1)$ on the pair $(R(S),\langle\cdot\,,
\cdot\rangle_H)\,$.}

\vspace{1ex}

Theorem 3.9(i) and corollary 3.11 above are extensions of results of \break
Schwarz, theorems 1$'$ and 2$'$ in \cite{S}, derived for the restricted case
where the cohomology of the resolvent vanishes. The derivations of both of
these results in \cite{S} use elliptic operator theory (heat kernel
expansion), however the preceding shows that they are related by the
straightforward algebraic result theorem 3.10.

Note that the formula (\ref{3.33}) is a generalisation to the present
infinite-dimensional case of the formula (\ref{2.40})
for the finite-dimensional case.

\section{Examples and Applications}

\subsection{A general example of a quadratic functional with elliptic
resolvent}

We now present a class of topological
quadratic functionals of the type discussed in \S3. Each of
these functionals has a canonical topological elliptic resolvent.
The corresponding partition functions
include the partition functions appearing in the semiclassical approximation
(\ref{1.14}) for the
Witten invariant. We will apply
the method and results of \S3 to evaluate these partition functions.
As a byproduct we will show that the metric-dependence of the Ray-Singer
torsion factors out in a simple way as a power of the volume of $M$ in
certain cases where the cohomology is non-vanishing.

Throughout the following we assume that the dimension
$n$ of the manifold $M$ (introduced in \S3) is odd.
Let $\chi\,\colon\,\pi_1(M){\rightarrow}O(V,\langle\cdot\,,\cdot\rangle_V)$
be a representation of $\pi_1(M)$
on a real vectorspace $V$ orthogonal w.r.t. an inner product
$\langle\cdot\,,\cdot\rangle_V$ in $V\,$. The by a standard construction in
differential geometry $\chi$ determines a real flat vectorbundle $\xi$
over $M$ and a flat connection map $\nabla$ on the space
$\Omega(M,\xi)$ of differential forms on $M$ with values in $\xi\,$
(i.e. the space of smooth
sections in the vectorbundle $\Lambda(TM)^*\otimes\xi\,$).
Let $\nabla_q$ denote the restriction of
$\nabla$ to the space $\Omega^q(M,\xi)$ of $q$-forms
and define the cohomology spaces
$$
H^q(\nabla)=ker(\nabla_q)\,\Big/\,Im(\nabla_{q-1})\,.
$$
The bundle $\xi$ has a canonical Hermitian structure
(determined by $\langle\cdot\,,\cdot\rangle_V\,$)
which we denote by $\lbrace\langle\cdot\,,\cdot
\rangle_x\rbrace_{x{\in}M}\,$ which $\nabla$ is compatible with.
This Hermitian structure determines for each $x{\in}M$
a linear map $\langle\,\cdot\,\rangle_x\,
\colon\,\xi_x\otimes\xi_x{\longrightarrow}R\,$,
which in turn determines linear maps
\begin{eqnarray*}
\Bigl({\Lambda}^p(T_xM)^*\otimes\xi_x\Bigr)
\otimes\Bigl({\Lambda}^q(T_xM)^*\otimes\xi_x\Bigr)
&\stackrel{\wedge}{\longrightarrow}&
{\Lambda}^{p+q}(T_xM)^*\otimes\Bigl(\xi_x\otimes\xi_x\Bigr)
\stackrel{\langle\,\cdot\,\rangle_x}{\longrightarrow} \\
&\longrightarrow&{\Lambda}^{p+q}(T_xM)^*\,.
\end{eqnarray*}
We denote the image of $\omega_x\otimes\tau_x$
under this map by $\langle\omega_x\wedge\tau_x
\rangle_x\,$.

Set $n=dimM$ and define $m$ by $n=2m+1\,$.
Assume to begin with that $m$ is
odd; in this case we define the
quadratic functional $S_{\nabla}$ on $\Omega^m(M,\xi)$ by
\begin{equation}
S_{\nabla}(\omega)=\int_M\langle\omega(x)\wedge(\nabla_{\!m}\omega)(x)
\rangle_x \label{4.1}
\end{equation}
where $\nabla_m$ is the restriction of $\nabla$ to $\Omega^m(M,\xi)\,$.
Choosing a metric on $M$
we can construct from the metric and Hermitian structure in $\xi$ a
Hermitian structure in $\Lambda(T_xM)^*\otimes\xi\,$ and from this
we get inner products $\langle\cdot\,,\cdot\rangle_q$
in the spaces $\Omega^q(M,\xi)$ by analogy with
(\ref{3.1}). This enables us to write
\begin{equation}
S_{\nabla}(\omega)=
\langle\omega,T_{\nabla}\omega\rangle_m\;\;\;\;,\;\;\;\;T_{\nabla}=*\nabla_m
\label{4.2}
\end{equation}
where $*$ is the Hodge-star map.
The map $T_{\nabla}$ is formally selfadjoint with the property
\begin{equation}
T_{\nabla}^2=\nabla_m^*\nabla_m\,. \label{4.3}
\end{equation}
There is a canonical topological elliptic resolvent
$R(S_{\nabla})$ for the functional (\ref{4.1}):
\begin{equation}
0\stackrel{0}{\longrightarrow}\Omega^0(M,\xi)
\stackrel{\nabla_0}{\longrightarrow}\dots
\stackrel{\nabla_{m-2}}{\longrightarrow}
\Omega^{m-1}(M,\xi)\stackrel{\nabla_{m-1}}
{\longrightarrow}ker(S_{\nabla})\stackrel{0}{\longrightarrow}0 \label{4.4}
\end{equation}
Comparing (\ref{4.4}) with (\ref{2.5})
we see that for the resolvent $R(S_{\nabla})$ we have
$N=m\,,\,\,
\Gamma_{\!k}=\Omega^{m-k}(M,\xi)\,,\,\,T_k=\nabla_{m-k}\,\,\mbox{and}
\,\,H^k(R(S_{\nabla}))=H^{m-k}(\nabla)\,$.
Note that by our previously established notation convention
$ker(S_{\nabla})\equiv
ker(T_{\nabla})=ker(\nabla_m)\,$.

Now choose an inner product
$\langle\cdot\,,\cdot\rangle_{H^k}$ in each space
$H^k(R(S_{\nabla}))\,$, then from
(\ref{3.4}) we find that the partition function
of $S_{\nabla}$ with the resolvent (\ref{4.4}) is :
\begin{equation}
Z(\beta\,;R(S_{\nabla}),\langle\cdot\,,\cdot\rangle_H,
\langle\cdot\,,\cdot\rangle)
=\pi^{\zeta/2}\,e^{-\frac{i\pi}{4}((\frac{2\theta}{\pi}\mp1)\zeta\,\pm\,
\eta)}\,
|\beta|^{-\zeta/2}\,\tau(M,\chi,\langle\cdot\,,\cdot\rangle_H)^{1/2}
\label{4.5}
\end{equation}
where we define for $m$ both odd and even :
\begin{equation}
\tau(M,\chi,\langle\cdot\,,\cdot\rangle_H)
=\widetilde{\tau}(M,\chi,g)\cdot\Psi
(M,\chi,g,\langle\cdot\,,\cdot\rangle_H)^2 \label{4.6}
\end{equation}
with
\begin{equation}
\widetilde{\tau}(M,\chi,g)=
\prod_{q=0}^{m-1}det(\widetilde{\nabla}_q^*
\widetilde{\nabla}_q)^{(-1)^q}
det(\widetilde{\nabla}_m^*\widetilde{\nabla}_m)^{\frac{1}{2}(-1)^m} \label{4.7}
\end{equation}
and
\begin{equation}
\Psi(M,\chi,g,\langle\cdot\,,\cdot\rangle_H)
=\prod_{q=0}^mdet(\phi_q^*\phi_q)^
{\frac{1}{2}(-1)^q} \label{4.8}
\end{equation}
where $\phi_q=\Phi_{m-q}\,\colon\,ker(\Delta_q)\stackrel{\simeq}
{\longrightarrow}H^q(\nabla)$ is the restriction of the projection
map $ker(\nabla_q){\to}H^q(\nabla)$ to $ker(\Delta_q)\,$.
In particular, for $\beta=i\lambda\,$, $\lambda{\in}{\bf R}_{\pm}\,$,
we have $\theta=\pm\pi/2$ and
\begin{equation}
Z(i{\lambda}\,;R(S_{\nabla}),\langle\cdot\,,\cdot\rangle_H,
\langle\cdot\,,\cdot\rangle)
=\pi^{\zeta/2}\,e^{\mp\frac{i\pi}{4}\eta}
\,k^{-\zeta/2}\,\tau(M,\chi,\langle\cdot\,,\cdot\rangle_H)^{1/2}\,. \label{4.9}
\end{equation}
We will discuss the quantities $\tau(M,\chi,\langle\cdot\,,\cdot\rangle
_H)\,,\;\zeta\,$, and $\eta$ appearing in the above expressions below.
First we give a treatment similar to the preceding for the case where
$m$ is even. In this case the functional (\ref{4.1}) is identically zero;
to see this note that
$$
\langle\omega(x)\wedge(\nabla_{\!m}\omega)(x)\rangle_x
=\langle(\nabla_{\!m}\omega)(x)\wedge\omega(x)\rangle_x
$$
which gives
\begin{eqnarray*}
d\langle\omega(x)\wedge\omega(x)\rangle_x
&=&\langle(\nabla_{\!m}\omega)(x)\wedge\omega(x)\rangle_x
+(-1)^m\langle\omega(x)\wedge(\nabla_{\!m}\omega)(x)\rangle_x \\
&=&(1+(-1)^m)\,\langle\omega(x)\wedge(\nabla_{\!m}\omega)(x)\rangle_x
\end{eqnarray*}
and the integral of this over $M$ vanishes by Stokes' theorem.
However, when $m$ is even it is
possible to define a non-vanishing
functional similar to (\ref{4.1}), the partition function of which is
given by an expression similar to
(\ref{4.5}): In the preceding we replace $\xi$ by the
complex vectorbundle $\xi\otimes{\bf C}\,$ and define the real-valued quadratic
functional $S_{\nabla}^{{\bf C}}(\omega)$ on $\Omega^m(M,\xi\otimes{\bf C})$ by
$$
S_{\nabla}^{{\bf C}}=i\int_M\langle\overline
{\omega(x)}\wedge(\nabla_{\!m}\omega)(x)
\rangle_x\,.
$$
This can be written as
\begin{equation}
S_{\nabla}^{{\bf C}}(\omega)=
\langle\omega,T_{\nabla}\omega\rangle_m\;\;\;,\;\;\;T_{\nabla}=i*\nabla_m
\label{4.9a}
\end{equation}
where $T_{\nabla}$ is formally selfadjoint and has the same property
(\ref{4.3}) as in the case where
$m$ is odd: $T_{\nabla}^2=\nabla_m^*\nabla_m\,$.
Replacing $\xi$ by $\xi\otimes{\bf C}$ in (\ref{4.4})
we obtain an elliptic resolvent $R(S_{\nabla}^{{\bf C}})$ for
$S_{\nabla}^{{\bf C}}\,$, and from it we obtain the
following expression for the partition function
of $S_{\nabla}^{{\bf C}}\,$ (c.f.
the remarks at the end of \S2) :
\begin{equation}
Z(\beta\,;R(S_{\nabla}^{{\bf C}}),\langle\cdot\,,\cdot\rangle_H
,\langle\cdot\,,\cdot\rangle)=\pi^{\zeta}\,
e^{-\frac{i\pi}{2}((\frac{2\theta}{\pi}\mp1)\zeta\,\pm\,\eta)}\,|\beta|^
{-\zeta}
\tau(M,\chi,\langle\cdot\,,\cdot\rangle_H)^{-1} \label{4.10}
\end{equation}
with $\tau(M,\chi,\langle\cdot\,,\cdot\rangle_H)$ defined by (\ref{4.6}).

In \cite{S} the quadratic
functional on $\Omega(M,\xi)=\oplus_{q=0}^n\Omega^q(M,\xi)$ defined by
replacing $\nabla_m$ by $\nabla=\oplus_{q=0}^n\nabla_q$ in
(\ref{4.1}) was considered. It was
shown to have a canonical topological
elliptic resolvent and the modulus of its partition function
was shown in the case where the cohomology
vanishes to be proportional to a power of the
Ray-Singer torsion of $\nabla\,$.
The resolvent is somewhat more involved than the resolvent
(\ref{4.4}) and we will not describe it here. The partition function of this
functional
can be evaluated by our methods;
the resulting expression is equal to
(\ref{4.5}) for $m$ odd and the square root of
(\ref{4.10}) for $m$ even.

Examples of other quadratic functionals with elliptic resolvents were given
in \cite{S}. Our methods and results also apply for these. However, they are
of a less general nature than the functional (\ref{4.1}) above (since their
construction requires choosing differential forms on the manifold) and
we will not discuss them here.

The quantity $\zeta$ appearing in the partition functions above
can be expressed in terms of the dimensions of the cohomology
spaces of $\nabla\,$:
Since $H^k(R(S_{\nabla}))=H^{m-k}(\nabla)\,$ for the resolvent (\ref{4.4})
it follows
from (\ref{3.25}) with $N=m$ that
\begin{equation}
\zeta=(-1)^{m+1}\sum_{q=0}^m(-1)^qdimH^q(\nabla)\,. \label{4.11}
\end{equation}
We now discuss the quantities (i) $\tau(M,\chi,\langle\cdot\,,\cdot\rangle_H)$
and (ii) $\eta$ appearing in the partition functions above.

(i) It follows from theorem 3.9 that $\tau(M,\chi,\langle\cdot\,,\cdot\rangle
_H)$ is independent of the choice of metric $g$ on $M\,$. In fact the
quantity $\widetilde{\tau}(M,\chi,g)$ defined in
(\ref{4.7}) is precisely the
Ray-Singer torsion \cite{RS1} of the representation $\chi$ of $\pi_1(M)$
constructed using the metric $g\,$, and $\tau(M,\chi,\langle\cdot\,,\cdot
\rangle_H)$ is a version of the Ray-Singer torsion as a ``function of the
cohomology'' defined and shown to be metric-independent in \cite[\S3]{RS2}.
By an argument analogous to that following
(\ref{2.26}) we see that given a choice
of orientation in each space $H^q(\nabla)\,,\;\,q=0,1,\dots,m$ then
considered as a functional of $\langle\cdot\,,\cdot\rangle_H$ we can
interpret $\tau(M,\chi,\langle\cdot\,,\cdot\rangle_H)^{1/2}$ as an element
\begin{equation}
\tau^{1/2}(M,\chi)\in\otimes_{q=0}^m\Lambda^{max}H^q(\nabla)^{*^q}\,.
\label{4.11a}
\end{equation}

We will now show that there is a canonical choice of (metric-independent)
inner product $\langle\cdot\,,\cdot\rangle_{H^0(\nabla)}$ in $H^0(\nabla)$
and use this to calculate $\tau(M,\chi,\langle\cdot\,,\cdot\rangle_H)$
in the case where $H^0(\nabla){\ne}0$ and $H^q(\nabla)=0$ for
$q=1,\dots,m\,$. The result shows that in this case the metric-dependence
of the Ray-Singer torsion $\widetilde{\tau}(M,\chi,g)$ factors out in a
simple way as a power of the volume of $M\,$.
The space $H^0(\nabla)$ is $ker(\nabla_0)\,$, i.e. the kernel of $\nabla$
restricted to the space of sections in $\xi\,$, and can be described as
follows. Choose a point $x_0{\in}M$ then $H^0(\nabla)$ can be identified
with the subspace $W\subseteq\xi_{x_0}$ having the property that the
parallel transport of any $w{\in}W$ to any fibre $\xi_x$ is independent of
the choice of curve from $x_0$ to $x\,$. (For simplicity we have assumed
that $M$ is connected; the generalisation to the case where $M$ has more
than one connected component is obvious). The inner product
$\langle\cdot\,,\cdot\rangle_{x_0}$ in $\xi_{x_0}$ induces an inner product
$\langle\cdot\,,\cdot\rangle_W$ in $W\,$, thereby determining an inner
product in $H^0(\nabla){\simeq}W\,$. Since $ker(\nabla_0)=H^0(\nabla)$ the
map $\phi_0\,\colon\,ker(\nabla_0)\stackrel{\simeq}{\longrightarrow}
H^0(\nabla)$ in (\ref{4.8}) is the identity map. However, since $ker(\nabla_0)$
and $H^0(\nabla)$ are equipped with different inner products the map
$\phi_0^*$ is not the identity. To determine $\phi_0^*$ we identify
$H^0(\nabla)$ with $W$ so that $\phi_0$ is the map $\phi_0\,\colon\,
ker(\nabla_0)\stackrel{\simeq}{\longrightarrow}W$ defined by
$$
\phi_0^{-1}(w)(x)=\mbox{parallel transport of $w$ from $\xi_{x_0}$
to $\xi_x$ along any curve.}
$$
Since $\nabla$ is compatible with the Hermitian structure in $\xi$
we have
$$
\langle\phi_0^{-1}(w)(x),\phi_0^{-1}(v)(x)\rangle_x=
{\langle}w,v\rangle_{x_0}={\langle}w,v\rangle_W
$$
for all $w\,,\,v{\in}W$
and $x{\in}M\,$, and it follows that
\begin{eqnarray*}
\langle\phi_0^{-1}(w),\phi_0^{-1}(v)\rangle_{\Omega^0(M,\xi)}
&=&\int_M\langle\phi_0^{-1}(w)(x),\phi_0^{-1}(v)(x)\rangle_xvol(x) \\
&=&{\langle}w,v\rangle_W\,V(M)
\end{eqnarray*}
where $V(M)$ is the volume of $M$ determined by the metric $g\,$. It
follows that $(\phi_0^{-1})^*\phi_0^{-1}=V(M){\cdot}I\,$, and therefore
$det(\phi_0^*\phi_0)^{-1}=V(M)^{dimH^0(\nabla)}\,$. It now follows from
(\ref{4.6}) and (\ref{4.8}) that for the case we are considering,
\begin{equation}
\tau(M,\chi,\langle\cdot\,,\cdot\rangle_H)=\widetilde{\tau}(M,\chi,g)
{\cdot}V(M)^{-dimH^0(\nabla)}\,. \label{4.11b}
\end{equation}
This shows the following

\vspace{1ex}

\noindent {\bf Theorem 4.1.} {\it
Assume $H^0(\nabla){\ne}0$ and $H^q(\nabla)=0$ for $q=1,\dots,m\,$
(where $n=2m+1$ is the dimension of $M\,$). Then the product
$$
\widetilde{\tau}(M,\chi,g){\cdot}V(M)^{-dimH^0(\nabla)}
$$
is independent of the choice of metric $g\,$, i.e. the metric-dependence
of the Ray-Singer torsion $\widetilde{\tau}(M,\chi,g)$ factors out as
$V(M)^{dimH^0(\nabla)}\,$. }

\vspace{1ex}

\noindent In the special case where $dimM=3\,,\,\;H^1(M)=0$ and $\chi$
is the trivial representation the metric-independence of (\ref{4.11b}),
i.e. of $\widetilde{\tau}(M,g){\cdot}V(M)^{-1}\,$, was pointed out by
Schwarz in \cite{S}.

(ii) We now show how
the dependence of $\eta=\eta(0\,|\,T_{\nabla})$ on the connection map
$\nabla$ can be
expressed via the formula \cite[(I)]{APS}
for the index of the twisted signature
operator for a certain vectorbundle over $M\times[0,1]\,$.
Fix an arbitrary flat connection map $\nabla^{(0)}$ on $\Omega(M,\xi)\,$,
define $\nabla^{(t)}=\nabla^{(0)}+t(\nabla-\nabla^{(0)})\;,\;
t{\in}[0,1]\,$, let $\widetilde{\xi}$
denote the pull-back of $\xi$ to a vectorbundle over
$M\times[0,1]$ and let $\widetilde{\nabla}$
denote the connection map on $\Omega(M\times[0,1],
\widetilde{\xi})$ determined by $\nabla^{(t)}\,$.
Pick a metric on $M$ and equip $[0,1]$ with the
natural metric and orientation; these induce
the product metric and orientation on $M\times[0,1]\,$.
Then, as in \cite[(I),\S4]{APS} we have the twisted signature map
$$
(\widetilde{\nabla}+\widetilde{\nabla}^*)_+\,:
\,\Omega_+(M\times[0,1],\xi\otimes{\bf C})
\longrightarrow\Omega_-(M\times[0,1],\xi\otimes{\bf C})\,.
$$
With the Atiyah-Patodi-Singer boundary conditions the index of this map
can be calculated from \cite[(I),theorem(3$\cdot$10)]{APS} to be
\begin{eqnarray}
I_+(\nabla,\nabla^{(0)})&=&\sum_{4j+2k=n+1}2^k\int_MLj(TM)
{\wedge}Tch_k(\nabla,\nabla^{(0)})
-\eta(0\,|\,T_{\nabla})\nonumber \\
& &\;\;+\eta(0\,|\,T_{\nabla^{(0)}})-\sum_{p=0}^m{\dim}H^p(\nabla)
-\sum_{q=0}^m{\dim}H^q(\nabla^{(0)})\,.\nonumber \\
& &\; \label{4.12}
\end{eqnarray}
Here $L_j(TM)$ is the j'th term in the Hirzebruch $L-$polynomial and
$$
Tch_k(\nabla,\nabla^{(0)})=\int_0^1(\frac{i}{2\pi})^k\frac{1}{(k-1)!}
Tr\Bigl((\nabla-\nabla^{(0)})\Omega_t^{k-1}\Bigr)dt
$$
is the transgression of the k'th Chern character (with $\Omega_t$ the
curvature tensor of $\nabla^{(t)}\,$). In deriving
(\ref{4.12}) we have used the
following formulae for the boundary operators $(-1)^lB^{(l)}$ of
$(\widetilde{\nabla}+\widetilde{\nabla}^*)_+\;,\;l=0,1\,$,
where $B^{(0)}$ is the
boundary operator at the boundary $M\times{\lbrace}0\rbrace$ of
$M\times[0,1]\,$, and $-B^{(1)}$ is the boundary operator at
$M\times{\lbrace}1\rbrace\,$. Setting $\nabla^{(1)}\equiv\nabla\,$ the
$B^{(l)}$ are elliptic selfadjoint maps on $\Omega(M,\xi)$ defined on
$q-$forms by\footnote{The expression for the boundary operator depends on
the convention used to define the Hodge star-map $*\,$.
To get (\ref{4.13}) we
use $*\alpha\wedge\beta=\langle\alpha,\beta{\rangle}vol\,$ (rather than
$\beta{\wedge}*\alpha=\langle\beta,\alpha{\rangle}vol\,$).
Then (\ref{4.13})
is in agreement with the expression \cite[(I),(4$\cdot$6)]{APS} derived in
the case where $m$ is odd.}
\begin{equation}
B_q^{(l)}=(-i)^{\lambda(q)}\,(*\nabla^{(l)}+(-1)^{q+1}\nabla^{(l)}*)
\label{4.13}
\end{equation}
where $\lambda(q)=(q+1)(q+2)+m+1\,$.
Note that $B^{(l)}$ maps even forms to even forms and odd forms to odd forms
and can therefore be decomposed as $B^{(l)}=B_{ev}^{(l)}{\oplus}
B_{odd}^{(l)}\,$. As pointed out in \cite[(I),\S4]{APS} $B_{ev}^{(l)}$
is isomorphic to $B_{odd}^{(l)}\,$, so
\begin{equation}
\eta(s\,|\,B_{ev}^{(l)})=\eta(s\,|\,B_{odd}^{(l)})
=\frac{1}{2}\eta(s\,|\,B^{(l)})\,. \label{4.14}
\end{equation}
The eta-functions of $B^{(l)}$ and $T_{\nabla^{(l)}}$ are now related by
combining (\ref{4.14}) with the following formulae:
\begin{eqnarray}
\eta(s\,|\,B_{odd}^{(l)})&=&\eta(s\,|\,*\nabla_m^{(l)})\quad\mbox{for $m$
odd}\label{4.15} \\
\eta(s\,|\,B_{ev}^{(l)})&=&\eta(s\,|\,i*\nabla_m^{(l)})\quad\mbox{for $m$
even}\label{4.16}
\end{eqnarray}
These are analogous to \cite[(I),proposition(4$\cdot$20)]{APS} and can be
derived by arguments analogous to the one given there. (They can also be
derived by arguments similar to the one used in the proof of theorem 3.3
above). Combining (\ref{4.15}),
(\ref{4.16}) and (\ref{4.14}) with the expressions (\ref{4.2})
and (\ref{4.9a}) for $T_{\nabla}$ we see that for $m$ even or odd,
\begin{equation}
\eta(s\,|\,B^{(l)})=2\eta(s\,|\,T_{\nabla^{(l)}})\,. \label{4.17}
\end{equation}
Finally, in deriving (\ref{4.12}) we have used the fact that
$(B^{(l)})^2=\oplus_{q=0}^n\Delta_q^{(l)}\,$ so that from Hodge-theory,
$$
dim\,kerB^{(l)}=\sum_{q=0}^ndimH^q(\nabla^{(l)})\,.
$$

Taking the index formula (\ref{4.12}) as an equation for
$\eta=\eta(0\,|\,T_{\nabla})$ we see that the metric-dependence
of $\eta$ enters through $L_j(TM)$
and $\eta(0\,|\,T_{\nabla^{(0)}})\,$. If $n=3$ then the only
contribution of the $L$-polynomial to
(\ref{4.12}) is $L_0=1$ and the metric-dependence of $\eta$ is
determined alone by $\eta(0\,|\,T_{\nabla^{(0)}})\,$.
Defining $B^{(t)}$ by replacing $\nabla^{(l)}$ by
$\nabla^{(t)}$ in (\ref{4.13})
the index $I_+(\nabla,\nabla^{(0)})$ is the spectral flow of $B^{(t)}\,$
(or equivalently, twice the spectral flow of $T_{\nabla^{(t)}}\,$) as $t$
runs from 0 to 1, see \cite[(III),\S\S7--8]{APS}.

\subsection{The partition functions in the semiclassical approximation
for the Witten-invariant}

We take the gauge group to be $G=SU(N)$ and define an inner product in the
Lie algebra ${\bf g}$ by ${\langle}a,b\rangle_{\bf g}=-\lambda_{\bf g}
\mbox{Tr}(ab)$ with scaling parameter $\lambda_{\bf g}>0\,$.
($SU(N)$ and its Lie algebra are identified with their fundamental
representations). The partition functions in the semiclassical
approximation (\ref{1.14}) are
partition functions of functionals of the form
$i{\lambda}S_{A_f}\,$, with $S_{A_f}=S_{\nabla}$ given by (\ref{4.1}) with
$m=1\,,\;\xi=M\times{\bf g}$ with Hermitian structure determined by
$\langle\cdot\,,\cdot\rangle_{\bf g}$ and $\nabla=d_{A_f}=d+ad(A_f)\,$
(where $A_f$ is a flat gauge field and $ad\,\colon\,{\bf g}{\rightarrow}
End({\bf g})$ is the adjoint representation, so $d_{A_f}\omega=d\omega
+[A_f,\omega]\,$), and $\lambda=
\frac{k}{4\pi\lambda_{\bf g}}\,$. These partition functions can be evaluated
by the method described in this paper; they are given by (\ref{4.9}) to be
(for $k>0\,$) :
\begin{equation}
(2\pi\sqrt{\lambda_{\bf g}})^{\zeta(A_f)}\,e^{-\frac{i\pi}{4}\eta(A_f,g)}\,
k^{-\zeta(A_f)/2}\,\tau(M,A_f,\langle\cdot\,,\cdot\rangle_{H(A_f)})^
{1/2} \label{4.22}
\end{equation}
where, from (\ref{4.11})
\begin{equation}
\zeta(A_f)=dimH^0(A_f)-dimH^1(A_f) \label{4.23}
\end{equation}
and from (\ref{4.2})
\begin{equation}
\eta(A_f,g)=\eta(0\,|\,T_{A_f})=\eta(0\,|\,*d_{A_f(1)}) \label{4.24}
\end{equation}
(this depends on a metric $g$ on $M$ through the Hodge star-map $*\,$).
We have substituted $A_f$ in the notation in place of $\nabla=d_{A_f}$
and in place of
the representation $\chi$ of $\pi_1(M)$ corresponding to $d_{A_f}\,$.
The partition function (\ref{4.22}) depends on a choice of inner products
$\langle\cdot\,,\cdot\rangle_{H^0(A_f)}$ and $\langle\cdot\,,\cdot\rangle_
{H^1(A_f)}$ in $H^0(A_f)$ and $H^1(A_f)$ respectively, which we denote
collectively by $\langle\cdot\,,\cdot\rangle_{H(A_f)}\,$.
{}From (\ref{4.11a}) we see that given a
choice of orientation in $H^0(A_f)$ and
$H^1(A_f)$ then as a functional of
$\langle\cdot\,,\cdot\rangle_{H(A_f)}$
we can interpret (\ref{4.22}) as an element in
$\Lambda^{max}H^0(A_f)\otimes\Lambda^{max}H^1(A_f)^*\,$.
Since the inner products $\langle\cdot\,,\cdot\rangle_q$ in the
$\Omega^q(M,{\bf g})$ are all proportional to $\lambda_{\bf g}$ the adjoint
maps $\nabla_{\!q}^*$ are independent of $\lambda_{\bf g}\,$. Therefore the
dependence of $\tau(M,A_f,\langle\cdot\,,\cdot\rangle_{H(A_f)})^{1/2}$ on
$\lambda_{\bf g}$ can only enter through the adjoint maps $\phi_q^*$
which appear in the factor
$det(\phi_0^*\phi_0)^{1/2}det(\phi_1^*\phi_1)^{-1/2}\,$ (c.f. (\ref{4.6}) and
(\ref{4.8})). If the inner products $\langle\cdot\,,\cdot\rangle_{H(A_f)}$
are chosen independent of $\lambda_{\bf g}$ then the partition function
(\ref{4.22}) is independent of $\lambda_{\bf g}\,$: In this case since
$\langle\cdot\,,\cdot\rangle_q\sim\lambda_{\bf g}$ we have
$\phi_q^*\sim\lambda_{\bf g}^{-1}$ so
$$
det(\phi_0^*\phi_0)^{1/2}det(\phi_1^*\phi_1)^{-1/2}\,\sim\,
\lambda_{\bf g}^{(-\frac{1}{2}dimH^0(A_f)+\frac{1}{2}dimH^1(A_f))}
$$
and this cancels the factor
$\sqrt{\lambda_{\bf g}}^{\zeta(A_f)}$ in (\ref{4.22}).
However, the canonical inner product in $H^0(A_f)$ described in the previous
subsection is proportional to $\lambda_{\bf g}$ in the present case.
Using it to define $\langle\cdot\,,\cdot\rangle_{H(A_f)}$ in cases where
$H^1(A_f)=0$ leads to $\tau(M,A_f,\langle\cdot\,,\cdot\rangle_{H(A_f)})$
being independent of $\lambda_{\bf g}\,$, so the dependence of the
partition function (\ref{4.22}) on $\lambda_{\bf g}$ in this case is given
by $\sqrt{\lambda_{\bf g}}^{\zeta(A_f)}=(\lambda_{\bf g})^{\frac{1}{2}
dimH^0(A_f)}\,$.

We now compare (\ref{4.22}) with the expressions previously calculated by
E.~Witten and conjectured by D.~Freed and R.~Gompf, and by L.~Jeffrey.
We will not include the geometric counterterm, which, as discussed in
the introduction, must be put in by hand to cancel the metric-dependence
of the phase.
The partition function of $S_{A_f}$ with $\beta=i$
was evaluated by Witten \cite[\S2]{W}
in the case where the cohomology of $d_{A_f}$ vanishes. This was done
using standard methods of quantum field theory, i.e. gauge fixing
implemented via a Lagrange-multiplier field. The method does not apply
when the cohomology is non-vanishing. The expression obtained was
\begin{equation}
e^{-\frac{i\pi}{4}\eta_w(A_f,g)}\,\widetilde{\tau}(M,A_f)^{1/2} \label{4.25}
\end{equation}
where $\widetilde{\tau}$ is the usual Ray-Singer torsion of $A_f$ and
$\eta_w(A_f,g)=\eta(0\,|\,B_{odd}^{(1)})$ with $B_{odd}^{(1)}$
the restriction of $B^{(1)}$ in (\ref{4.13}) to odd-degree forms
with $m=1$ and $\nabla^{(1)}=d_{A_f}\,$. From (\ref{4.14}), (\ref{4.15})
and (\ref{4.17})
we have $\eta_w(A_f,g)=\eta(A_f,g)\,$, so
(\ref{4.25}) is identical to (\ref{4.22})
when the cohomology of $d_{A_f}$ vanishes.

Based on (\ref{4.25}) Freed and Gompf \cite[\S1]{FG}
conjectured an expression for the partition function of the functional
$i\frac{k}{4\pi}S_{A_f}$ appearing in the semiclassical approximation
(\ref{1.14}) for the general case where the cohomology of $d_{A_f}$ is not
required to vanish. Some refinements were added by L.~Jeffrey in
\cite[\S5]{J} leading to the following expression:
\begin{equation}
\frac{1}{|C(G)|}e^{-\frac{i\pi}{4}\eta(A_f,g)}\,k^{-\frac{1}{2}(dimH^0(A_f)
-dimH^1(A_f))}\,<\tau^{1/2}(M,A_f),v_0>_0\,. \label{4.26}
\end{equation}
The motivation for the new features of this expression relative to (\ref{4.25})
is as follows: In \cite{FG} the $k-$dependence of the Witten-invariant
in the large$-k$ limit was calculated numerically for a number of
lens spaces and Brieskorn spheres and found to be
given by
\begin{equation}
k^{\stackrel{max}{A_f}\lbrace-\frac{1}{2}dimH^0(A_f)+
\frac{1}{2}dimH^1(A_f)\rbrace} \label{4.27}
\end{equation}
This led Freed and Gompf to conjecture the $k-$dependence
\begin{equation}
k^{-\frac{1}{2}(dimH^0(A_f)-dimH^1(A_f))} \label{4.28}
\end{equation}
in (\ref{4.26})\footnote{According to \cite{FG}
(\ref{4.28}) was originally suggested
by Witten in an informal lecture.}.

The phase $e^{-\frac{i\pi}{4}\eta(A_f,g)}$ in (\ref{4.26}) was also shown
numerically in \cite{FG} to be in agreement with the phase of the
Witten-invariant in the large$-k$ limit (provided the geometric
counterterms are included). The phase factor $\eta(A_f,g)$ was expressed
in \cite[(1.31)]{FG} via the index
formula for the self-dual complex over $M\times[0,1]$ determined by
$d_{A_f}\,$; it is also obtained as a special case of (\ref{4.12}) with
$\nabla^{(0)}=d\,$.
(To see
this note that the index $I_{sd}(A_f)$ of the self-dual complex
\cite[(1.30)]{FG} is half $I_+(d_{A_f},d)\,$, and for $G=SU(2)$
the transgression of the Chern character $Tch(d_{A_f},d)$ is
$CS(ad(A_f))=4CS(A_f)\,$ (identifying $su(2)$ with its fundamental
representation) where $CS(\cdot)$ is the Chern-Simons functional (\ref{1.13}),
and $dimH^q(d)=3dimH^q(M)\,$).

The factor $<\tau^{1/2}(M,A_f),v_0>_0$ in (\ref{4.26}) is due to L. Jeffrey
(in the case where $H^1(A_f){\ne}0\,$) and is defined as follows. As
previously noted, given a choice of orientation of $H^0(A_f)$ and
$H^1(A_f)$ we can consider $\tau^{1/2}$ as an element in $\Lambda^{max}
H^0(A_f)\otimes\Lambda^{max}H^1(A_f)^*\,$. Given a volume form
$v_0\in\Lambda^{max}H^0(A_f)^*\,$, we obtain $<\tau^{1/2}(M,A_f),v_0>_0
\in\Lambda^{max}H^1(A_f)^*\,$, where $<\cdot,\cdot>_0$ denotes the natural
pairing of $\Lambda^{max}H^0(A_f)$ and $\Lambda^{max}H^0(A_f)^*\,$.
In certain cases
$H^1(A_f)$ can be identified with the tangentspace of the moduli space of
flat gauge fields at $A_f\,$, then
$<\tau^{1/2}(M,A_f),v_0>_0$ is a top-degree
form on this moduli space. It follows that for these cases
the expression (\ref{1.14}) for the
semiclassical approximation can be generalised to the case where the moduli
space of flat gauge fields is not discrete:
the sum in (\ref{1.14}) can be replaced
by an integral over the moduli space. Note that if $v_0$ is determined by
the inner product $\langle\cdot\,,\cdot\rangle_{H^0(A_f)}$ in $H^0(A_f)$
then $<\tau^{1/2}(M,A_f),v_0>_0$ in (\ref{4.26}) can be identified with
$\tau(M,A_f,\langle\cdot\,,\cdot\rangle_{H(A_f)})^{1/2}$ in (\ref{4.22})
considered as a functional of the inner product $\langle\cdot\,,\cdot\rangle
_{H^1(A_f)}$ in $H^1(A_f)\,$. Finally, $|C(G)|$ in
(\ref{4.26}) is the order of
the center $C(G)$ of $G\,$; it is included to take account of the fact that
constant gauge transformations corresponding to elements in $C(G)$ act
trivially on gauge fields.

Comparing (\ref{4.22}) with (\ref{4.26}) we see that
(\ref{4.22}) considered as a functional
of the inner product $\langle\cdot\,,\cdot\rangle_{H^1(A_f)}$ in
$H^1(A_f)$ agrees with (\ref{4.26}) up to a numerical factor
$(2\pi)^{\zeta(A_f)}|C(G)|\,$. In particular, the phases and
$k-$dependencies  of (\ref{4.22}) and (\ref{4.26}) agree
(although recall that we have omitted the geometric counterterm in the
phase of the conjectured expression).
As described in the introduction, for $G=SU(2)$
the conjectured formula (\ref{4.26}) has been
shown numerically \cite{FG} and analytically \cite{J}, \cite{Roz}
to be in agreement
with the large$-k$ limit of the Witten-invariant for a wide class of
3-manifolds (lens spaces, Seifert manifolds and some Brieskorn spheres)
up to minor numerical factors. (These typically involve a $\sqrt{2}\,$,
although for $S^3$ a factor $\sqrt{\pi}$ is also involved).
Thus, for these 3-manifolds, the semiclassical
approximation with the partition functions calculated by the method
described in this paper to give (\ref{4.22}), with the geometric
counterterms added by hand to the phase,
agrees with the large$-k$
limit of the Witten-invariant modulo some numerical factors.

We now give an explanation for the discrepancy in the numerical factors
between the semiclassical approximation and the large$-k$ limit of the
Witten-invariant. It can be understood from recent work by the first author
\cite{Ad} combined with results of L.~Rozansky \cite{Roz}.
In \cite{Ad} a method for formally constructing the perturbative
expansion of the formal expression (\ref{1.12}) for the Witten-invariant
about a flat gauge field
$A_f$ is described. It is
based on a refinement of the Faddeev-Popov gauge-fixing procedure to
take account of the fact that the Lorentz gauge-fixing condition
$d_{A_f}^*\omega=0$ does not determine $\omega$ uniquely, since, as is
easily checked, the space of solutions $\omega$ is invariant under the
subgroup ${\cal G}_{A_f}$ of gauge transformations which act trivially
on $A_f\,$. The method requires $H^1(A_f)=0$ but does not require
$H^0(A_f)=0\,$; it is inspired by, and extends the work of S. Axelrod
and I. Singer \cite{AxSi} which requires $H^0(A_f)=H^1(A_f)=0\,$.
A generally assumed feature of perturbative QFT is that the term in the
semiclassical approximation corresponding to a solution $A_f$ of the
classical field equations is the same as the lowest order term in a
perturbative expansion about $A_f\,$. However, in \cite{Ad} we find
that the lowest order term differs from the term in the semiclassical
approximation by the inverse volume factor $V({\cal G}_{A_f})^{-1}\,$.
This factor also appeared in the work of Rozansky \cite{Roz}: He invoked the
normalisation procedure used to obtain physical quantities in QFT
to argue that each partition function in the semiclassical approximation
(\ref{1.14}) should be modified by dividing by $V({\cal G}_{A_f})\,$.
(In other words, the factor $V({\cal G}_{A_f})^{-1}$ is put in by hand in
\cite{Roz}, whereas in \cite{Ad} it is obtained by formal calculations
carried out inside a self-contained framework).
It is well-known that the isotropy groups ${\cal G}_{A_f}$ can be identified
with subgroups of $G\,$, so the volumes $V({\cal G}_{A_f})$ are finite.

Rozansky replaced the factor $|C(G)|$
in the conjectured expression (\ref{4.26})
with the factor
$V({\cal G}_{A_f})^{-1}\,$, and invoked ``an implicit factor $(\sqrt{2}\pi)
^{-1}$ coming with each of the 1-dimensional integrals'' in the integral
(\ref{1.1}) to obtain a modified conjectured expression for each partition
function in the semiclassical approximation (\ref{1.14}).
Up to a power of $\sqrt{2}$ this is the same as multiplying our
expression (\ref{4.22}) by
$V({\cal G}_{A_f})^{-1}\,$. Rozansky showed that with this expression the
semiclassical approximation agrees with the large$-k$ limit of the
Witten-invariant in all the cases for which this limit has been analytically
calculated (i.e. $S^3\,$, lens spaces, Seifert manifolds).
It follows that up to a power of $\sqrt{2}$
the lowest order term in the sum over $A_f$ of
the perturbative
expansions of (\ref{1.12}) about all non-equivalent
$A_f\,$, calculated by the method of \cite{Ad}, agrees with the large$-k$
limit of the Witten-invariant for these manifolds, as predicted by
QFT. Complete agreement can be obtained by modifying
the values of the volumes $V({\cal G}_{A_f})^{-1}$ in
\cite{Roz} by powers of $\sqrt{2}\,$.\footnote{There is sufficient
arbitrariness in the ``reasonable'' values for $V({\cal G}_{A_f})$
given in \cite{Roz} to allow for this.}
In particular this gives a self-contained method for deriving the factor
$\pi$ in the large$-k$ limit $\sqrt{2}{\pi}k^{-3/2}\,$
of the Witten-invariant for $S^3\,$,
which has previously
been a source of mystery\footnote{See e.g. \cite[p.113 problem (1)]{FG},
\cite[footnote p.587]{J}.}. We will demonstrate this explicitly below.

An intuitive explanation for why the partition functions in the
semiclassical approximation (\ref{1.14}), calculated by
the method described in this paper,
fail to produce the factors $V({\cal G}_{A_f})^{-1}$ is the following:
Loosely speaking, the semiclassical approximation (\ref{1.14}) does not know
about ${\cal G}_{A_f}\,$, it only knows about its infinitessimal version,
i.e. its Lie algebra, which is $H^0(A_f)\,$.
Therefore the factors $V({\cal G}_{A_f})^{-1}$ in the perturbative
expansions are replaced by the divergent factors $V(H^0(A_f))^{-1}$ in the
semiclassical approximation, which are divided out (i.e. discarded) when
normalising the partition functions. Indeed, the formal expression for
each partition function includes divergent volume factors
$V(\Gamma_{\!k})^{\pm1}$ and $V(H^k(R(S)))^{\pm1}\,$
(see (\ref{2.23})); when
$H^1(A_f)=0$ the contribution to these volumes from the cohomology
in the present case is
precisely $V(H^1(R(S)))^{-1}=V(H^0(A_f))^{-1}\,$.

We now give an illustration of the points discussed above by explicitly
calculating the semiclassical approximation
(\ref{1.14}) for the manifold $S^3$
with $G=SU(2)\,$.
Up to gauge equivalence $A_f=0$ is the only flat gauge field on $S^3\,$,
so the semiclassical approximation obtained from our extension of
Schwarz's method is given by (\ref{4.22}) with $A_f=0\,$. We include Witten's
geometric counterterm; in the present case this cancels precisely the
phase in (\ref{4.22}).
For reasons discussed above
we include the inverse volume factor $V({\cal G}_{A_f})^{-1}\,$.
For $A_f=0$ clearly ${\cal G}_{A_f}=G=SU(2)\,$, so this factor
is $V(SU(2))^{-1}\,$. Then, since $H^0(A_f)=su(2)$ and $H^1(A_f)=0$ we get
from (\ref{4.22})
\begin{equation}
(2\pi)^3\,\lambda_{\bf g}^{3/2}\,
k^{-3/2}\,\tau(S^3,A_f\!=\!0,\langle\cdot\,,\cdot
\rangle_{H^0(A_f=0)})
^{1/2}\,V(SU(2))^{-1} \label{4.29}
\end{equation}
where as before we take the inner product in
${\bf g}=su(2)$ to be ${\langle}a,b\rangle=-\lambda_{\bf g}
Tr(ab)\,$. This
gives an inner product in $H^0(A_f\!=\!0)\,{\cong}\,su(2)\,$ and determines
$V(SU(2))=(\frac{2}{\lambda_{\bf g}})^{3/2}16\pi^2\,$.
By the same calculations as those leading to (\ref{4.11b})  we find
\begin{equation}
\tau(S^3,A_f\!=\!0,\langle\cdot\,,\cdot\rangle_{H^0(A_f=0)})^{1/2}=
\widetilde{\tau}(S^3)^{3/2}\,V(S^3)^{-3/2} \label{4.30}
\end{equation}
where $\widetilde{\tau}(S^3)$ is the usual Ray-Singer torsion of $S^3\,$
and $V(S^3)$ is its volume.
The standard embedding $S^3\hookrightarrow{\bf R}^4$ induces a metric in
$S^3$ from the standard metric in ${\bf R^4}\,$; using this we get
$V(S^3)=2\pi^2\,$. From the calculations in \cite[\S4]{Ray} we get
$\widetilde{\tau}(S^3)=e^{-4\zeta_R'(0)}\,$, where $\zeta_R(s)$ is the
analytic continuation of the Riemann zeta-function.
{}From \cite[p.26]{Bateman}
$\,\zeta_R'(0)=-\frac{1}{2}\log(2\pi)\,$,
so $\widetilde{\tau}(S^3)=(2\pi)^2\,$.
Substituting these expressions into (\ref{4.29}) gives
\begin{equation}
\frac{(2\pi)^3\lambda_{\bf g}^{3/2}k^{-3/2}\widetilde{\tau}(S^3)^{3/2}}
{V(S^3)^{3/2}V(SU(2))}
=\frac{(2\pi)^3\lambda_{\bf g}^{3/2}k^{-3/2}(2\pi)^3}{(2\pi^2)^{3/2}(\frac{2}
{\lambda_{\bf g}})^{3/2}16\pi^2}
=\sqrt{2}{\pi}k^{-3/2}\,(\frac{\lambda_{\bf g}}{\sqrt{2}})^3\,.
\label{4.31}
\end{equation}
QFT predicts that this should coincide with the Witten-invariant
$Z_W(k)$ in the limit of large $k\,$. The calculation of the Witten-invariant
for $S^3\,$ \cite[\S4]{W} involves very different mathematics to that
used to obtain
(\ref{4.31}): $\,S^3$ is obtained by surgery on $S^2{\times}S^1$
and $Z_W(k)$ for $S^3$ turns out to be a matrix element of the representation
of the corresponding diffeomorphism of the torus on the characters of the
irreducible level$-k$ representations of the loop group of $G\,$. For
$G=SU(2)$ this is
\begin{equation}
Z_W(k)=
\sqrt{\frac{2}{k+2}}\,\sin\Bigl(\frac{\pi}{k+2}\Bigr)\;\sim\;
\sqrt{2}{\pi}k^{-3/2}\;\;\;\;\mbox{for}\,\;k\rightarrow\infty\,. \label{4.32}
\end{equation}
Thus (\ref{4.31}) reproduces the
factor $\pi$ in the large$-k$ limit (\ref{4.32}).
To reproduce the correct power of $\sqrt{2}$ in (\ref{4.31}) we must set
$\lambda_{\bf g}=\sqrt{2}\,$, a somewhat peculiar value.

\vspace{3ex}

\begin{center}

\noindent {\large\bf Appendix}

\end{center}

\vspace{1ex}

For $A=(A_N,\dots,A_1,A_0){\in}GL(\Gamma_{\!N})\times\cdots{\times}
GL(\Gamma_{\!1}){\times}GL(\Gamma)$ define $T_k^{*(A)}$ to be the
adjoint of $T_k\,\colon\,\Gamma_{\!k}\longrightarrow\Gamma_{\!k-1}$
determined by the inner products $\langle\cdot\,,\cdot\rangle_k^{A_k}$
and $\langle\cdot\,,\cdot\rangle_{k-1}^{A_{k-1}}\,$. Define $W^{\perp(A)}$
to be the orthogonal complement of a subspace $W\subseteq\Gamma_{\!k}$
determined by $\langle\cdot\,,\cdot\rangle_k^{A_k}\,$;
it is easy to see that $W^{\perp(A)}=(A_k^*A_k)^{-1}(W^{\perp})\,$.
In the following we will use the relations $T_k=T_k(P_k+Q_k)=\widetilde{T}Q_k$
and, when the cohomology of the resolvent vanishes,
$T_k=(P_{k-1}+Q_{k-1})T_k=P_{k-1}T_k\,$. Recall that the cohomology
of the resolvent is assumed to vanish for (\ref{2.13}) and (\ref{2.14}),
but not for (\ref{2.31}).

\vspace{1ex}

{\it Proof of formula (\ref{2.13}).}
\hfill\break
We calculate $det(\widetilde{T}_{A_0})\,$, where $\widetilde{T}_{A_0}$ is
the restriction of $T_{A_0}=(A_0^*A_0)^{-1}T$ to $ker(T_{A_0})^{\perp(A)}=
ker(T)^{\perp(A)}=(A_0^*A_0)^{-1}(ker(T)^{\perp})\,$:
\begin{eqnarray*}
det(\widetilde{T}_{A_0})&=&det\Bigl((A_0^*A_0)^{-1}T\Bigl|_{(A_0^*A_0)^{-1}
(ker(T)^{\perp})}\Bigr) \\
&=&det\Bigl(T(A_0^*A_0)^{-1}\Bigl|_{Im(T)=ker(T)^{\perp}}\Bigr) \\
&=&det\Bigl(\widetilde{T}Q_0(A_0^*A_0)^{-1}Q_0\Bigr) \\
&=&det(\widetilde{T})\,det(Q_0(A_0^*A_0)^{-1}Q_0)
\end{eqnarray*}
proving formula (\ref{2.13}).

\vspace{1ex}

{\it Proof of formula (\ref{2.14}).}
\hfill\break
We calculate $det((\widetilde{T}_{k(A)})^{*(A)}\widetilde{T}_{k(A)})\,$,
where $\widetilde{T}_{k(A)}$ is the restriction of $T_k$ to
$ker(T_k)^{\perp(A)}=(A_k^*A_k)^{-1}(ker(T_k))\,$.
For $v\in\Gamma_{\!k}\,,\;w\in\Gamma_{\!k-1}$ we have
$$
{\langle}w,T_kv\rangle_{k-1}^{A_{k-1}}=\langle(A_k^*A_k)^{-1}T_k^*
(A_{k-1}^*A_{k-1})w,v\rangle_k^{A_k}
$$
so $T_k^{*(A)}=(A_k^*A_k)^{-1}T_k^*(A_{k-1}^*A_{k-1})\,$. It follows that
\begin{eqnarray*}
det(\widetilde{T}_{k(A)}^*\widetilde{T}_{k(A)})
&=&det\Bigl(T_k^{*(A)}T_k\Bigl|_{(A_k^*A_k)^{-1}(ker(T_k)^{\perp})}\Bigr) \\
&=&det\Bigl((A_k^*A_k)^{-1}T_k^*(A_{k-1}^*A_{k-1})T_k\Bigl|_{(A_k^*A_k)
^{-1}(ker(T_k)^{\perp})}\Bigr) \\
&=&det\Bigl(T_k(A_k^*A_k)^{-1}T_k^*(A_{k-1}^*A_{k-1})\Bigl|_{Im(T_k)}
\Bigr) \\
&=&det\Bigl(\widetilde{T}_kQ_k(A_k^*A_k)^{-1}Q_k\widetilde{T}_k^*P_{k-1}
(A_{k-1}^*A_{k-1})P_{k-1}\Bigr) \\
&=&det(\widetilde{T}_k\widetilde{T}_k^*)\,det\Bigl(Q_k(A_k^*A_k)^{-1}Q_k\Bigr)
\\
& &\;\times\;det\Bigl(P_{k-1}(A_{k-1}^*A_{k-1})P_{k-1}\Bigr)
\end{eqnarray*}
proving formula (\ref{2.14}).

\vspace{1ex}

{\it Proof of formula (\ref{2.31}).}
\hfill\break
Define ${\cal H\/}_{k(A)}$ and $\Phi_{k(A)}\,\colon\,{\cal H\/}_{k(A)}
\stackrel{\simeq}{\longrightarrow}H^k(R(S))$ as in
(\ref{2.18}) and (\ref{2.19}) with
the inner product $\langle\cdot\,,\cdot\rangle_k$ in $\Gamma_{\!k}$ replaced
by $\langle\cdot\,,\cdot\rangle_k^{A_k}\,$.
Abbreviating $H^k(R(S))$ by $H^k$ we define maps for $k=1,\dots,N\,$:
\begin{eqnarray*}
{\cal \/T}_k
&=&T_k\oplus\Phi_{k-1}^{-1}\,\colon\,\Gamma_{\!k}{\oplus}H^{k-1}
{\longrightarrow}Im(T_k)\oplus{\cal H\/}_{k-1}=ker(T_{k-1}) \\
{\cal T\/}_{k(A)}
&=&T_k\oplus\Phi_{k-1(A)}^{-1}\,\colon\,\Gamma_{\!k}{\oplus}H^{k-1}
{\longrightarrow}Im(T_k)\oplus_{A_{k-1}}{\cal H\/}
_{k-1(A)}=ker(T_{k-1})
\end{eqnarray*}
where $\oplus_{A_{k-1}}$ denotes direct sum orthogonal w.r.t.
$\langle\cdot\,,\cdot\rangle_{k-1}^{A_{k-1}}\,$. Then the formula
(\ref{2.31}) is
equivalent to
\begin{eqnarray*}
\lefteqn{det\Bigl({\cal T\/}_{k(A)}^{*(A)}{\cal T\/}_{k(A)}
\Bigl|_{ker(T_k)^{\perp(A)}{\oplus}H^{k-1}}\Bigr)} \\
&=&det\Bigl(Q_k(A_k^*A_k)^{-1}Q_k\Bigr)\,
det\Bigl(P_{k-1}(A_{k-1}^*A_{k-1})
P_{k-1}\Bigr) \\
& &\;\times\;
det\Bigl({\cal T\/}_k^*{\cal T\/}_k\Bigl|_{ker(T_k)^{\perp}
{\oplus}H^{k-1}}\Bigr)\;\;\;\;\;\;\;\;\;\;\;\;\;\;\;\;\;\;\;
\;\;\;\;\;\;\;\;\;\;\;\;\;\;\;\;\;\;\;\;\;\;\;\;\;\;(A.1)
\end{eqnarray*}
where the adjoint map ${\cal T\/}_{k(A)}^{*(A)}$ is defined w.r.t. the
inner products $\langle\cdot\,,\cdot\rangle_k^{A_k}\oplus\langle\cdot\,,
\cdot\rangle_{H^{k-1}}$ in $\Gamma_{\!k}{\oplus}H^{k-1}$ and
$\langle\cdot\,,\cdot\rangle_{k-1}^{A_{k-1}}$ in $\Gamma_{\!k-1}\,$.
An argument analogous to the preceding proof of (\ref{2.14}) gives
\begin{eqnarray*}
\lefteqn{det\Bigl({\cal T\/}_{k(A)}^{*(A)}{\cal T\/}_{k(A)}\Bigl|
_{ker(T_k)^{\perp(A)}{\oplus}H^{k-1}}\Bigr)} \\
&=&det\Bigl(Q_k(A_k^*A_k)^{-1}Q_k\Bigr)\,
det\Bigl(P_{k-1}(A_{k-1}^*A_{k-1})P_{k-1}
\Bigr) \\
& &\;\times\;
det\Bigl({\cal T\/}_{k(A)}^*{\cal T\/}_{k(A)}\Bigl|_
{ker(T_k)^{\perp}{\oplus}H^{k-1}}\Bigr)\;\;\;\;\;\;\;\;\;\;\;\;\;\;\;\;\;
\;\;\;\;\;\;\;\;\;\;\;\;\;\;\;\;\;\;\;\;\;\;(A.2)
\end{eqnarray*}
It is clear from the definitions that $P_{[{\cal H\/}_{k-1}]}\Phi_{k-1(A)}
^{-1}=\Phi_{k-1}^{-1}\,$, where $P_{[{\cal H\/}_{k-1}]}$ is the orthogonal
projection map of $\Gamma_{\!k-1}$ onto ${\cal H\/}_{k-1}\,$. It follows
that the restriction of ${\cal T\/}_{k(A)}$
to $ker(T_k)^{\perp}{\oplus}H^{k-1}$
has the form
$$
\left(\begin{array}{cc}
\widetilde{T}_k & Q_k\Phi_{k-1(A)}^{-1} \\
0 & \Phi_{k-1}^{-1}
\end{array}\right)\,\colon\,ker(T_k)^{\perp}{\oplus}H^{k-1}\stackrel{\simeq}
{\longrightarrow}Im(T_k)\oplus{\cal H\/}_{k-1}
$$
so
\begin{eqnarray*}
det\Bigl({\cal T\/}_{k(A)}^*{\cal T\/}_{k(A)}\Bigl|_
{ker(T_k){\oplus}H^{k-1}}\Bigr)
&=&det(\widetilde{T}_k^*\widetilde{T}_k)\,det((\Phi_{k-1}^{-1})^*
\Phi_{k-1}^{-1}) \\
&=&det\Bigl({\cal T\/}_k^*{\cal T\/}_k\Bigl|_{ker(T_k)^{\perp}{\oplus}
H^{k-1}}\Bigr)\,.
\end{eqnarray*}
Substituting this into (A.2) gives (A.1), proving formula (\ref{2.31}).

\end{document}